\documentstyle[onecolumn]{mn}

\newcounter{parentequation}
\def\pmb#1{\setbox0=\hbox{#1}%
    \kern-.025em\copy\kern-\wd0
    \kern.05em\copy\kern-\wd0
    \kern-.025em\raise.0433em\box0}
\def\beglet{
  \addtocounter{equation}{1}%
  \setcounter{parentequation}{\value{equation}}%
  \setcounter{equation}{0}%
  \def\theequation{\arabic{parentequation}\alph{equation}}%
  \ignorespaces
}
\def\endlet{
  \setcounter{equation}{\value{parentequation}}%
  \def\theequation{\arabic{equation}}%
}

\def\ltsima{$\; \buildrel < \over \sim \;$}
\def\gtsima{$\; \buildrel > \over \sim \;$}
\def\simlt{\lower.5ex\hbox{\ltsima}}
\def\simgt{\lower.5ex\hbox{\gtsima}}
\def\p2Y{\;_2Y}
\def\m2Y{\;_{-2}Y}

\def\etal{{\it et al.}\rm}
\def\etals{{\it et al. }\rm}
\def\mk2{\mu {\rm K}^2}

\begin{document}

\title[Polarization Power Spectrum Estimation]
{Hybrid Estimation of CMB Polarization Power Spectra}

 \author[G. Efstathiou]{G. Efstathiou\\
Institute of Astronomy, Madingley Road, Cambridge, CB3 OHA.}

\maketitle

\begin{abstract}
This paper generalises the hybrid power spectrum estimator developed
in Efstathiou (2004a) to the estimation of polarization power spectra of
the cosmic microwave background radiation.  The hybrid power spectrum
estimator is unbiased and we show that it is close to optimal at all
multipoles, provided the pixel noise satisfies certain reasonable
constraints. Furthermore, the hybrid estimator is computationally fast
and can easily be incorporated in a Monte-Carlo chain for {\it
Planck}-sized data sets. Simple formulae are given for the covariance
matrices, including instrumental noise, and these are tested
extensively against numerical simulations. We compare the behaviour of
simple pseudo-$C_\ell$ estimates with maximum likelihood estimates at low
multipoles. For realistic sky cuts, maximum likelihood estimates
reduce very significantly the mixing of $E$ and $B$ modes. To achieve
limits on the scalar-tensor ratio of $r \ll 0.1$ from sky maps with
realistic sky cuts, maximum likelihood methods, or pseudo-$C_\ell$
estimators based on unambiguous $E$ and $B$ modes, will be essential.

\vskip 0.1 truein


\vskip 0.3 truein

\end{abstract}

\section{Introduction}

In an earlier paper (Efstathiou 2004a, hereafter E04) a hybrid power
spectrum estimator was developed for temperature anisotropies of the
cosmic microwave background (CMB) anisotropies. This estimator
combined quadratic maximum likelihood (QML) estimates at low
multipoles with a set of pseudo-$C_\ell$ (PCL) estimates  at higher
multipoles to produce a near optimal power spectrum estimate over the
entire multipole range for realistic sky coverages, scanning patterns
and instrument noise.  We used analytic arguments and large numbers of
numerical simulations to demonstrate that the method was near optimal
and that the covariance matrix over the full range of multipoles could
be estimated simply and accurately.

A full maximum-likelihood estimator would require the inversion and
multiplication of $N_d \times N_d$ matrices, where $N_d$ is the size
of the data vector.  For {\it WMAP} (Bennett \etals 2003a) or {\it
Planck} (Bersanelli \etals 1996; The Planck Consortia 2005)
sized data sets, with $N_d \simgt 10^6 -10^7$, a brute force
application of ${\cal O} (N_d^3)$ methods is impossible
computationally.  This has been portrayed as a major computational
challenge in CMB data analysis ({\it e.g.} Bond \etals 1999; Borrill
1999) and various approximate, but computationally demanding,
techniques for solving this problem have been proposed ({\it e.g.} Oh,
Spergel and Hinshaw 1999, Dor\'e, Knox and Peel (2001), Pen
20003). One of the major motivations for E04 was to show that an
${\cal O} (N_d^3)$ computation is unnecessary and that a near-optimal
estimator (essentially indistinguishable from an exact ${\cal O}
(N_d^3)$ maximum-likelihood solution), with a calculable covariance
matrix, could be estimated simply by combining PCL and QML
estimates. Furthermore, the analysis of realistic experiments must
deal with uncertainties such as correlated instrument noise, beam
calibrations, foreground separation, point sources {\it etc}.  It is
simply not worth performing an exact ${\cal O} (N_d^3)$ power spectrum
analysis if the assumptions under which it is optimal are violated by
`real world' complexities, and if these complexities cannot be folded
into the error estimates.  With a fast hybrid estimator it is feasible
to assess such `real-world' complexities using a hybrid estimator
within a Monte-Carlo chain and to quantify any corrrections to the
covariance matrix, or likelihood function. The ability to analyse such
`real-world' effects is likely to be much more important than any
hypothetical marginal improvement that might be gained by applying a
full ${\cal O} (N_d^3)$ method. We refer the reader to E04 for a
discussion of the rationale for a hybrid estimator. The arguments will
not be repeated here. Instead, we will focus on generalising E04 to
the estimation of power spectra of the CMB polarization anisotropies.

As is well-known, the Thomson scattering of an anisotropic photon
distribution leads to a small net linear polarization of the CMB
anisotropies (for an introductory review and references to earlier
work see Hu and White 1997). This polarization signal can be
decomposed into scalar $E$-modes and pseudo-scalar $B$-modes. The
separation of a polarization pattern into $E$ and $B$ modes is of
particular interest since scalar primordial perturbations generate
only $E$ mode while tensor perturbations generate $E$ and $B$ modes of
roughly comparable amplitudes (Zaldarriaga and Seljak 1997;
Kamionkowski, Kosowsky and Stebbins 1997). The detection of an
intrinsic $B$-mode signal in the CMB would provide incontrovertible
evidence that the Universe experienced an inflationary phase and would
fix the energy scale of inflation (see {\it e.g.} Lyth 1984).

An $E$-mode polarization signal was first discovered by DASI\footnote{
Degree Angular Scale Interferometer} (Kovac \etals 2002, Leitch \etals
2004).  Exquisite measurements of the temperature-$E$-mode (TE)
cross power spectrum have been reported by the WMAP\footnote{Wilkinson
Microwave Anisotropy Probe} team (Kogut \etals 2003). Measurements of
the $E$-mode power spectrum have been reported by the
CBI\footnote{Cosmic Background Imager} experiment (Readhead \etals
2004) and by the 2003 flight of Boomerang (Montroy \etals 2005).
Primordial $B$-mode anisotropies have not yet been detected in the
CMB. The detection of $E$-mode and possibly $B$-mode anistropies is one
of the main science goals of the {\it Planck} mission and of ground
based experiments such as {\it Clover} (Taylor \etals 2004).

Associated with these experimental developments, there have been many
investigations of techniques for analysing $E$ and $B$ modes from maps
of the CMB sky. These analyses can be grouped, approximately, into the
following catagories:

\begin{itemize}
\item[(i)] {\it PCL estimators and correlation functions:} Fast
methods of estimating $E$ and $B$-mode power spectra using correlation
functions are described by Chon \etals (2004). Statistically
equivalent PCL estimators are described by Kogut \etals (2003), Hansen
and Gorski (2003), Challinor and Chon (2005, hereafter CC05) and Brown
\etals (2005).  In particular, CC05 present analytic approximations to
the covariance matrices of PCL estimates for the case of noise-free
data, while Brown \etals (2005) develop a Monte-Carlo method for
calibrating covariance estimates from incomplete maps of the sky. PCL
estimators can be evaluated using fast spherical transforms and hence
scale as ${\cal O} (N_d^{3/2})$.

\item[(ii)] {\it Maximum Likelihood Methods:} The generalization of
the iterative maximum likelihood power spectrum estimation methods of
Bond \etals (1998) to the analysis of polarisation is straightforward
and will not be discussed further here. Tegmark and de Oliveira-Costa
(2001, hereafter TdO01) define a quadratic estimator which is based on
assumed forms for the temperature and polarization power spectra,
generalising earlier work of Tegmark (1997). This method, which we
will refer to as QML, is equivalent to a maximum likelihood solution
if the guesses for the power spectra are close to their true values.
As explained above, these methods involve matrix inversions and
multiplications which scale as ${\cal O} (N_d^{3})$.

\item[(iii)] {\it Harmonic $E$ and $B$ Mode Decomposition:} The
Stokes' parameters $Q$ and $U$ describing linear polarization define a
rank two symmetric trace-free polarization tensor on sphere. Over the
complete sky, the polarization tensor can be decomposed uniquely into
$E$ and $B$ modes. However, since the $E$ and $B$ decomposition is
non-local, it is non-unique in the presence of boundaries. In any
realistic situation, a sky cut must be imposed to exclude
contamination of the CMB signal by high levels of Galactic emission at
low Galactic latitudes. Various authors have discussed ways of
detecting pure $E$ and $B$ modes from $Q$ and $U$ maps on an
incomplete sky (Lewis \etals 2002; Bunn \etals 2003; Bunn 2003; Lewis
2003). A key motivation for these analyses has been for diagnostic
purposes ({\it e.g.} checking for a frequency dependent Galactic
$B$-mode signal). However, as this article was nearing completion an
interesting paper appeared by Smith (2005) describing how to construct
pseudo-$C_\ell$ estimators from unambiguous $E$ and $B$ modes on a cut
sky. This type of analysis can effectively eliminate the severe $E$
and $B$ mode mixing that afflicts simple pseudo-$C_\ell$ estimators at
low multipoles (see Section 3.4). (The method is not completely
straightforward, nor unique, because it requires a `pre-estimation' step
to define unambiguous modes. We will throughout this paper use the
abbreviation PCL to refer to the simple pseudo-$C_\ell$ estimators as
defined in Section 2.)

\end{itemize}

As we will show in this paper, these three analysis techniques are
closely related. To illustrate how these methods are interelated
it is useful to consider separately the cases of noise-free and
noisy data:

\vskip 0.05 truein

\noindent
$\bullet$ {\it Comparison of PCL and QML estimators for noise-free
data:} On a complete sky with noise-free data, PCL and QML estimates
are identical. However, a sky cut will couple CMB modes over a range
of multipoles $\Delta L$, and this will lead to mixing of $E$ and $B$
pseudo-multipoles defined over the incomplete sky (see Section
2). Although PCL power spectrum estimators can be defined which give
unbiased estimates of the true $E$ and $B$ mode power spectra {\it in
the mean}, the mixing of $E$ and $B$ modes is reflected in large
variances and couplings between the estimated power spectra at
multipoles $\ell \simlt \Delta L$ (Section 2). A QML estimator applied
to noise-free data unscrambles the $E$- and $B$-mode multipoles, in
much the same way as the direct modal decompositions described by
Lewis \etals (2002), Lewis (2003) and Smith (2005). The QML estimator
returns almost optimal power spectrum estimates with smaller variances
than a PCL estimator (Section 3). In particular, for realistic sky
cuts an estimator that minimises $E$ and $B$ mode mixing, such as the
QML estimator, is essential if one wants to probe low amplitude
$B$-mode signals (tensor-scalar ratios $r \ll 0.1$). At high
multipoles, $\ell \gg \Delta L$, $E$ and $B$ mode mixing becomes
unimportant and, in the case of noise free data, QML and PCL
estimators become statistically equivalent (Section 3).

\vskip 0.05 truein

\noindent
$\bullet$ {\it Comparison of PCL and QML estimators for noisy data:} If the
following conditions are satisfied: (a) the $Q$ and $U$ maps have
identical noise properties; (b) the noise is uncorrelated with
variance per pixel $\sigma^2_i$; (c) the estimates of $E$ and $B$ mode
power spectra are for multipoles higher than some characterstic
multipole $L_N$, where noise dominates; then one can show that an
inverse-variance weighted PCL estimator is statistically equivalent to
a QML estimator (Section 4.2). For the intermediate multipoles, $ L_N \simlt l
\simlt \Delta L$, the optimal weighting for PCL estimates is
intermediate between equal weight per pixel and inverse variance
weighting. By combining a set of PCL estimates with different weights,
it is possible to define a fast polarization estimator (in analogy
with the temperature estimator discussed by E04) that is statistically
indistinguishable from a maximum likelihood estimator over a wide
range of multipoles.  Dealing with strongly correlated noise is more
problematic. Fortunately, for {\it Planck}-type scanning strategies,
the noise pattern (in $Q$, $U$ and $T$) should be accurately white at
high multipoles (Efstathiou 2005).  A detailed correlated noise model
should therefore only be required at low multipoles. It is
straightforward to include a model for correlated noise in a QML
estimate at low multipoles, provided the pixel noise covariance matrix
can be estimated for low resolution maps (Section 4.2).

The layout of this paper is similar to that for the temperature
analysis presented in E04. PCL polarization estimates are discussed in
Section 2 together with analytic approximations for covariances
matrices in the absence of instrumental noise. The relationship of
this work to the results presented in CC05 is discussed. Section 3
discusses QML polarization estimators for noise free data. A large set
of numerical simulations is used to illustrate the effects of $E$ and
$B$ mode mixing on an incomplete sky for both PCL and QML
estimators. Section 4 discusses PCL and QML estimators including
instrumental noise and Section 5 discusses a hybrid polarization power
spectrum estimator. To keep the discussion simple, most of the
analytic and numerical results presented in this paper refer to $E$
and $B$ mode power spectrum estimation. Our discussion of the
temperature-$E$ mode cross power spectrum (denoted $C^X$ in this
paper) is, intentionally, less complete since no new concepts are
required for its analysis. Our conclusions are summarized in Section
6.

\section{Estimation using Pseudo-$C_\ell$}

\begin{figure*}

\vskip 4.6 truein

\includegraphics{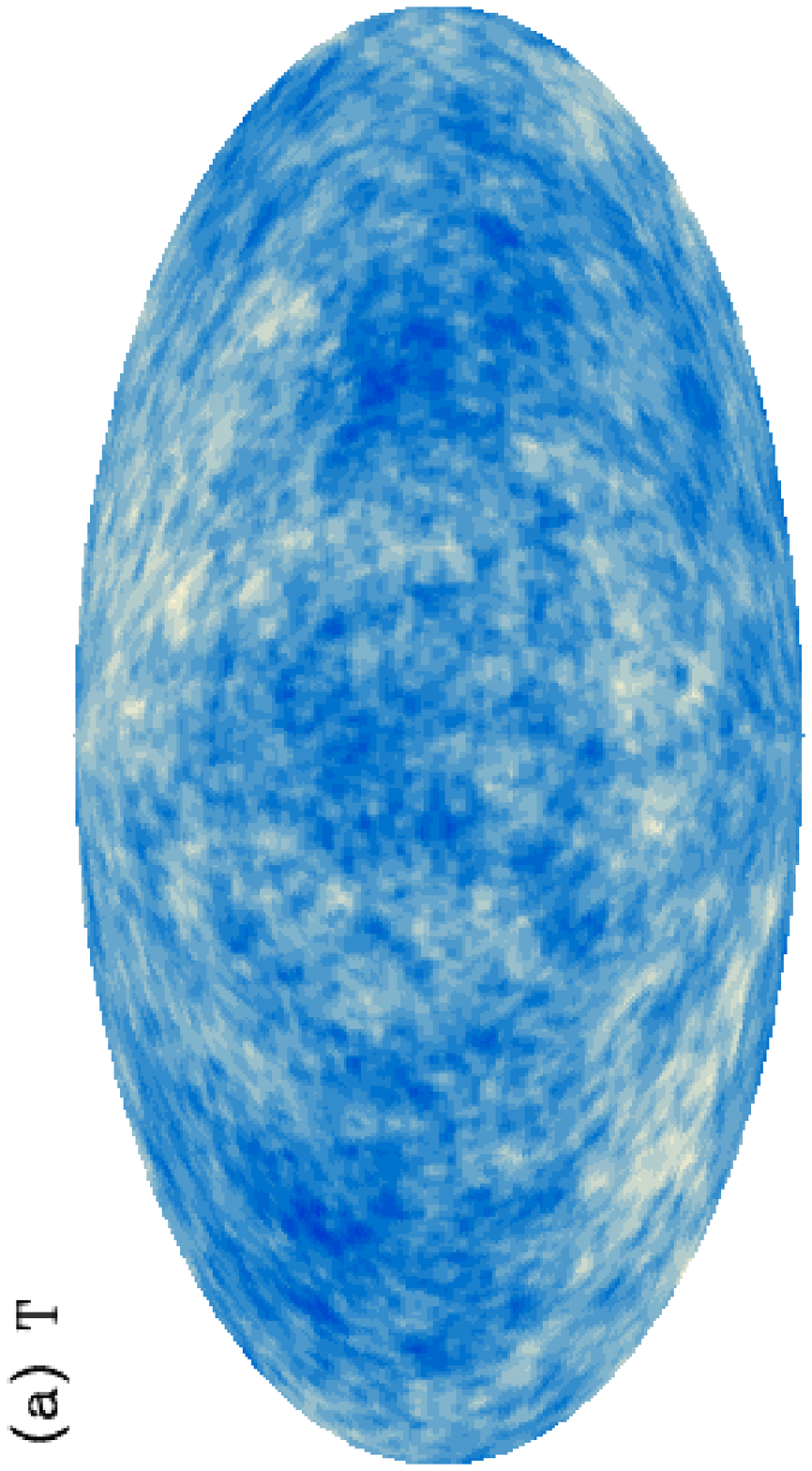}
\includegraphics{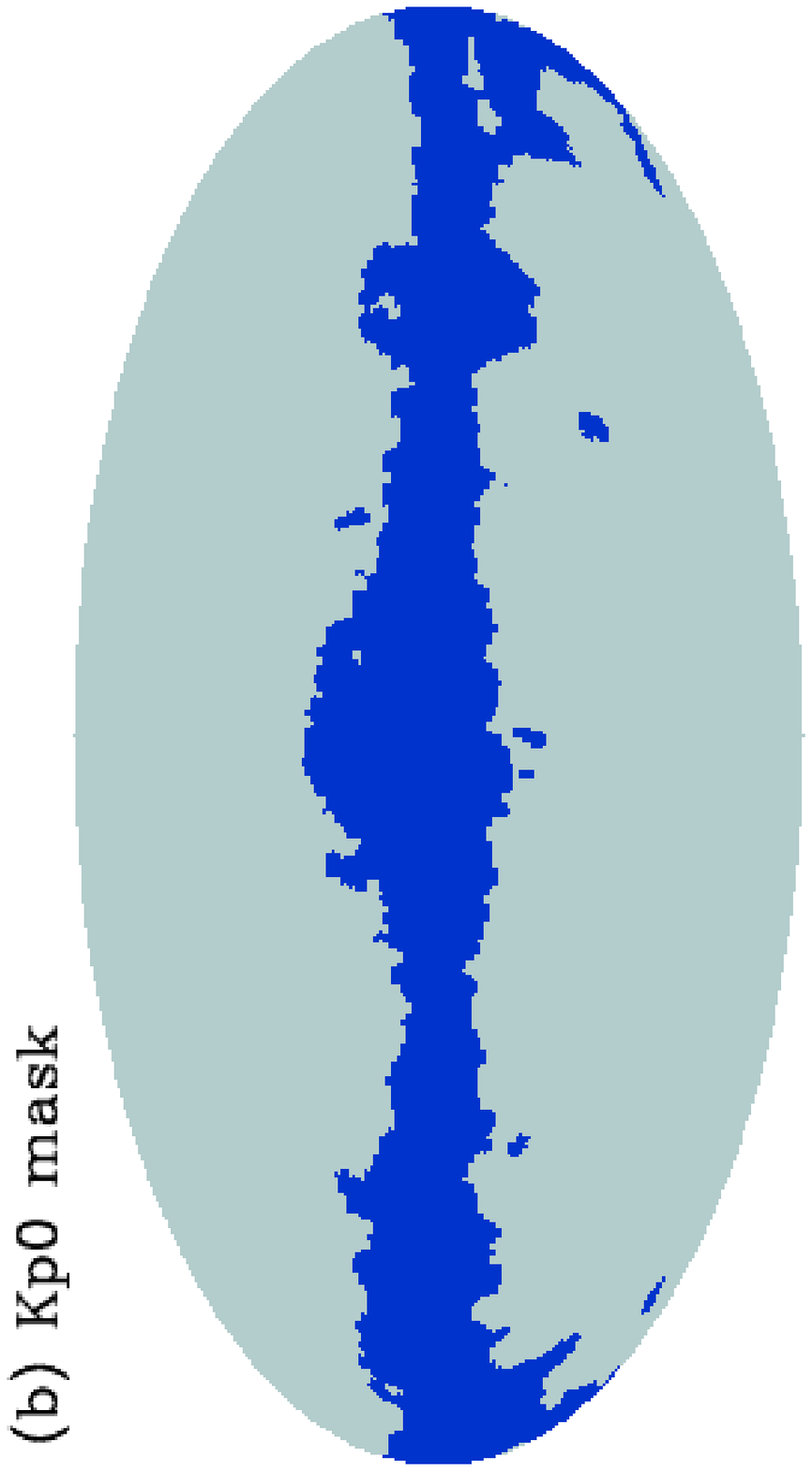}
\includegraphics{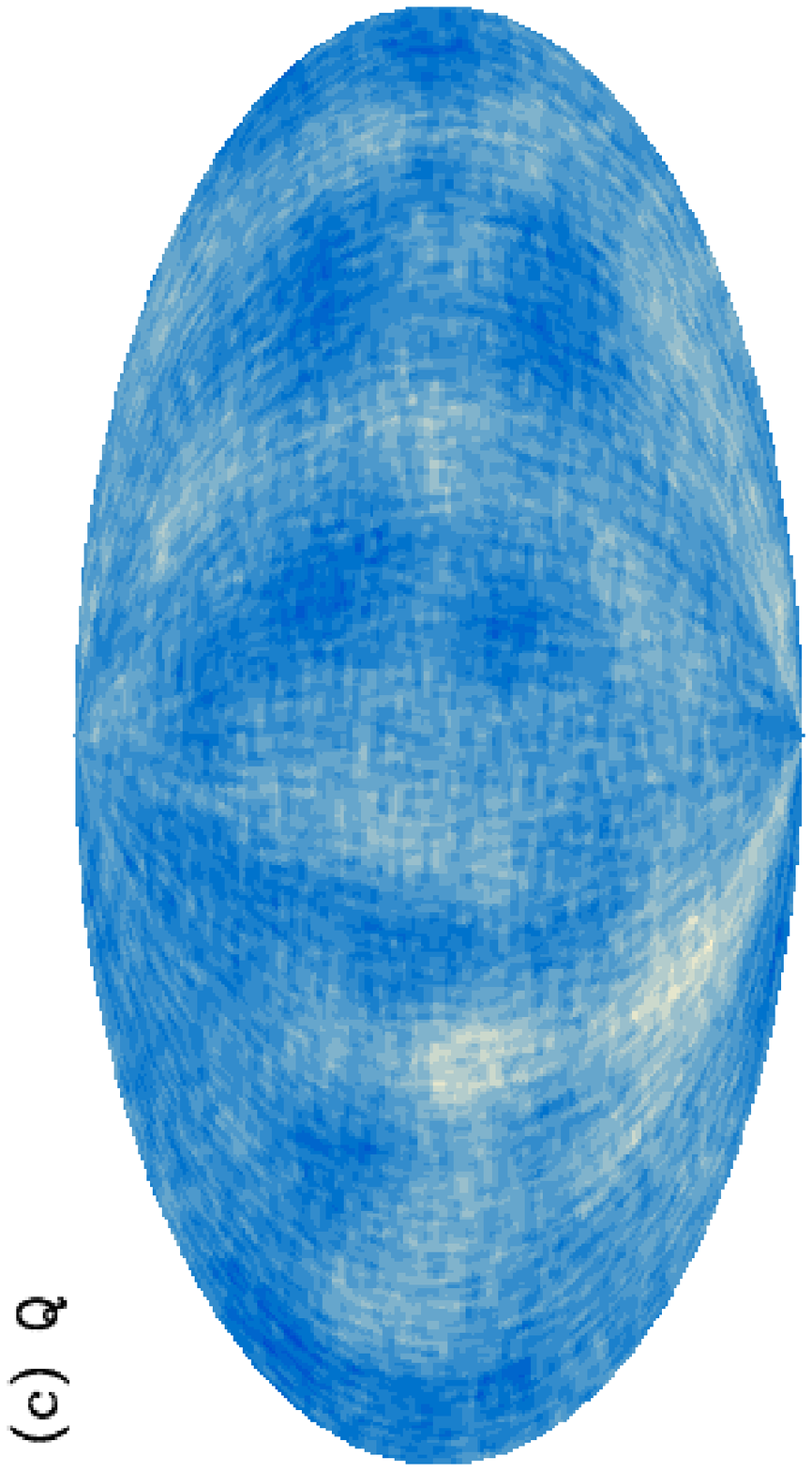}
\includegraphics{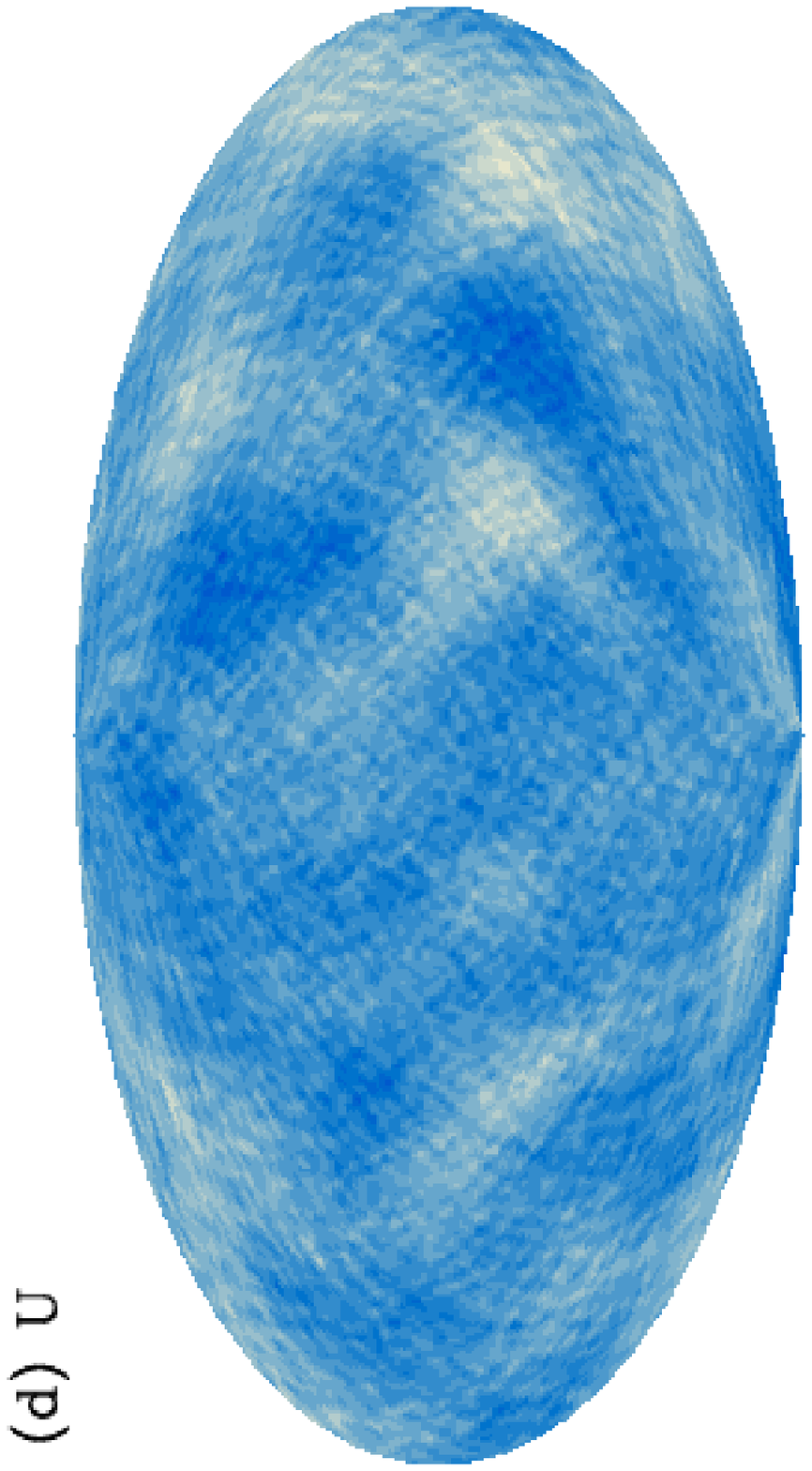}

\caption
{An example of the input maps used in the simulations described in Section 2.
The pixel size is $\theta_c = 1^\circ$ and the CMB sky has been smoothed
with a Gaussian beam of FWHM $\theta_s = 2^\circ$. The concordance $\Lambda$CDM
model with the parameters given in the text has been assumed and a
tensor component with  $r= 0.2$ has been added to give an non-zero
B-mode contribution. Figures (a), (c) and (d) show the $T$, $Q$, and
$U$ maps, while Figure (b) shows the WMAP Kp0 mask.} 

\label{figure1}

\end{figure*}

We begin by relating the $Q$ and $U$ Stokes parameters (defined in direction
${\bf \hat n}$ with respect to a spherical coordinate system ${\bf \hat e}_\theta$,
${\bf \hat e_\phi}$) and the spin two harmonics:
\begin{equation}
 Q({\bf \hat n})  \pm i U ({\bf \hat n}) = \sum_{\ell m} a_{\pm 2\ell m} \;_{\pm 2}Y_{\ell m}
({\bf \hat n}).
 \label{PCL1}
\end{equation}
Note that the sign convention for polarization in this paper follows that of Zaldarriaga and Seljak (1997),
which differs from the IAU convention (see Hamaker and Bregman 1996). The coefficients $a_{\pm 2\ell m}$ are
related to the $E$ and $B$ mode multipole coefficients $a^E_{\ell m}$ and $a^B_{\ell m}$ by
\begin{equation}
 a_{\pm 2 \ell m} = - (a^E_{\ell m}  \pm i a^B_{\ell m}).    \label{PCL2}
\end{equation}
If we have an incomplete sky, we can compute `pseudo-multipole' coefficients $\tilde a^T_{\ell m}$,
$\tilde a^E_{\ell m}$ and $\tilde a^B_{\ell m}$ by computing the following sums,
\beglet
\begin{equation}
 \tilde a^T_{\ell m} = \sum_i \Delta T_i w_i 
\Omega_i Y^*_{\ell m}(\theta_i),    \label{PCL3a}
\end{equation}
\begin{equation}
 \tilde a^E_{\ell m} = -{1 \over 2}\sum_i (Q+iU)_i w_i \Omega_i \p2Y^*_{\ell m}
+ (Q-iU)_i w_i \Omega_i \m2Y^*_{\ell m} = -{1 \over 2} \sum_i (Q_i R^{+*}_{\ell m} + 
iU_iR^{-*}_{\ell m}) w_i \Omega_i,    \label{PCL3b}
\end{equation}
\begin{equation}
 \tilde a^B_{\ell m} = \;\;{i \over 2}\sum_i (Q+iU)_i w_i \Omega_i\p2Y^*_{\ell m}
- (Q-iU)_i w_i \Omega_i\m2Y^*_{\ell m} =  \;\;{i \over 2} \sum_i (Q_i R^{-*}_{\ell m} + 
iU_iR^{+*}_{\ell m}) w_i \Omega_i,    \label{PCL3c}
\end{equation}
\endlet
where in equations (\ref{PCL3b}) and (\ref{PCL3c})
\begin{equation}
  R^{+}_{\ell m} = \p2Y_{\ell m} + \m2Y_{\ell m}, \qquad R^{-}_{\ell m} =
  \p2Y_{\ell m} - \m2Y_{\ell m}, \label{PCL3d}
\end{equation}
$\Omega_i$ is the area of pixel $i$ and $w_i$ is an arbitrary weight function with 
spherical transform 
\begin{equation}
 \tilde w_{\ell m} = \sum  w_i 
\Omega_i Y^*_{\ell m}(\theta_i).    \label{PCL3e}
\end{equation}
From these pseudo-multipole coefficients, we can form the following
PCL power spectrum estimates
\begin{equation}
   \tilde C^T_\ell = {1 \over (2 \ell + 1)} \sum_m \tilde a^T_{\ell m}\tilde
a^{T*}_{\ell m} , \quad   \tilde C^X_\ell = 
{1 \over (2 \ell + 1)} \sum_m \tilde a^T_{\ell m}\tilde
a^{E*}_{\ell m} ,  \quad \tilde C^E_\ell = 
{1 \over (2 \ell + 1)} \sum_m \tilde a^E_{\ell m}\tilde
a^{E*}_{\ell m}, \quad \quad \tilde C^B_\ell = 
{1 \over (2 \ell + 1)} \sum_m \tilde a^B_{\ell m}\tilde
a^{B*}_{\ell m}.
\label{PCL4}
\end{equation}
The expectation values of these PCL estimates are related to the true
values $C^T$, $C^X$, $C^E$, $C^B$ by
\begin{equation}
{\left ( \begin{array}{c} \langle \tilde C^T \rangle \\
\langle \tilde C^X \rangle \\
\langle \tilde C^E \rangle \\
\langle \tilde C^B \rangle \end{array} \right ) }  = 
{\left ( \begin{array}{cccc}  
 M^T & 0  & 0  & 0  \\
  0  & M^{X} & 0 & 0 \\
  0  & 0 & M^{EE} & M^{EB} \\
  0  & 0 & M^{BE} & M^{BB} \end{array} \right ) } 
{\left ( \begin{array}{c}  C^T \\
 C^X \\
 C^E \\
 C^B \end{array} \right ) },  \label{PCL5}
\end{equation}
where the matrices $M^T$, $M^X$, {\it etc} are given by (Kogut \etals 2003)
\beglet
\begin{equation}
  M^T_{\ell_1 \ell_2}  =  {(2 \ell_2 + 1) \over 4 \pi}
\sum_{\ell_3 }  (2 \ell_3 + 1)\tilde W _{\ell_3}
{\left ( \begin{array}{ccc}
        \ell_1 & \ell_2 & \ell_3  \\
        0  & 0 & 0
       \end{array} \right )^2}, \label{PCL6a}
\end{equation}
\begin{equation}
  M^X_{\ell_1 \ell_2}  =  
{(2 \ell_2 + 1) \over 8 \pi}
\sum_{\ell_3 } (2 \ell_3 + 1) \tilde W _{\ell_3}
(1 + (-1)^L)
{\left ( \begin{array}{ccc}
        \ell_1 & \ell_2 & \ell_3  \\
        0  & 0 & 0
       \end{array} \right ) } {\left ( \begin{array}{ccc}
        \ell_1 & \ell_2 & \ell_3  \\
        -2  & 2 & 0
       \end{array} \right ) }, \qquad L = \ell_1 + \ell_2 + \ell_3, \label{PCL6b}
\end{equation}
\begin{equation}
  M^{EE}_{\ell_1 \ell_2}  =    M^{BB}_{\ell_1 \ell_2} = 
{(2 \ell_2 + 1) \over 16 \pi}
\sum_{\ell_3 }  (2 \ell_3 + 1) \tilde W _{\ell_3}
(1 + (-1)^L)^2
 {\left ( \begin{array}{ccc}
        \ell_1 & \ell_2 & \ell_3  \\
        -2  & 2 & 0
       \end{array} \right )^2 },  \label{PCL6c}
\end{equation}
\begin{equation}
  M^{EB}_{\ell_1 \ell_2}  =    M^{BE}_{\ell_1 \ell_2} = 
{(2 \ell_2 + 1) \over 16 \pi}
\sum_{\ell_3 }  (2 \ell_3 + 1) \tilde W _{\ell_3}
(1 - (-1)^L)^2
 {\left ( \begin{array}{ccc}
        \ell_1 & \ell_2 & \ell_3  \\
        -2  & 2 & 0
       \end{array} \right )^2 }, \label{PCL6d}
\end{equation}
\endlet
and $\tilde W_\ell$ is the power spectrum of the weight function
\begin{equation}
   \tilde W_\ell = {1 \over (2 \ell + 1)} \sum_m \vert \tilde w_{\ell m} \vert ^2.  \label{PCL7}
\end{equation}
({\it cf} equation (\ref {PCL3e})).  In the absence of parity
violating physics in the early universe, the primordial $C^{TB}$ and
$C^{EB}$ spectra should be identically zero. Even if there is no
compelling motivation from fundamental physics, there may be other
reasons for wanting to estimate these spectra, {\it e.g.} for
consistency checks and to test for systematic errors.  However, to
keep the analysis as simple as possible, they will be not be included
in this paper.

If the sky cut is small, then the matrix $M$ appearing in equation 
(\ref{PCL5}) will be non-singular and hence invertible. If this is 
the case, then unbiased estimates of the true power spectra can
be formed from the PCL estimates by evaluating
\footnote{The notation here follows the notation introduced in E04.
The expectation values of the hatted PCL estimates ($\hat C_\ell$) (if
they are well defined) are equal to the true power spectra. The
expectation values of the tilde estimates ($\tilde C_\ell$) are given
by convolutions of the true power spectra (equation \ref{PCL5}).}
\begin{equation}
   \hat C_\ell = M^{-1}_{\ell \ell^{\prime}} \tilde C_{\ell^\prime}. 
 \label{PCL8}
\end{equation}

\begin{figure*}

\vskip 3.7 truein

\includegraphics{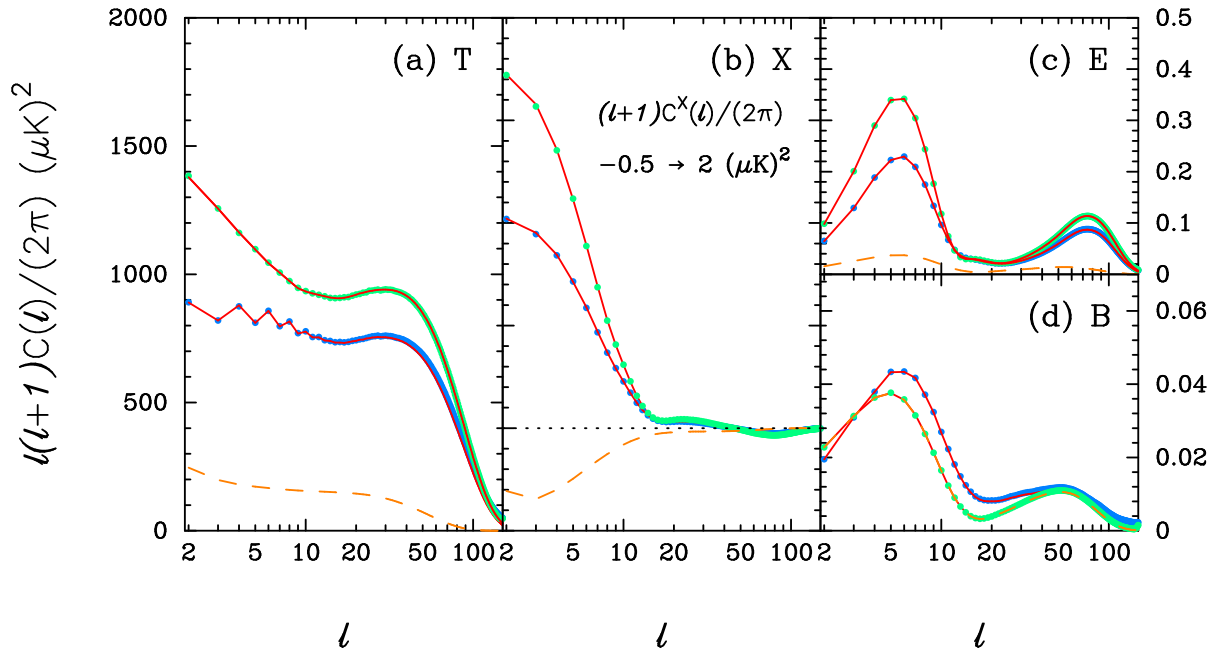}

\caption
{PCL polarization power spectra. The blue points show the convolved
power spectra, $\tilde C_\ell$, averaged over $10^5$ simulations with
the same parameters as those in Figure 1 and including the Kp0
mask. The green points show the deconvolved spectra $\hat C_\ell$. The
solid lines show the forms of $\tilde C_\ell$ and $\hat C_\ell$
computed from the theoretical input spectra. The dashed lines in
Figures (a)--(c) show the contribution to $\hat C_\ell$ from the
tensor component.}

\label{figure2}

\end{figure*}

As an example, Figure 1 shows noise-free $T$, $Q$ and $U$ maps from a
single Gaussian realisation of the concordance $\Lambda$CDM model
favoured by WMAP (Spergel \etals 2003).  (The exact parameters adopted
in this paper are those of the $\Lambda$CDM model defined in Section 2
of Efstathiou (2003)). The maps have been created using software
written by the author that uses an `igloo' pixelisation scheme. The
maps shown in Figure 1 have a pixel size of $1^\circ$ and a
symmetrical Gaussian beam smoothing of FWMH of $\theta_s =
2^\circ$. In these simulations, we have assumed a tensor-scalar ratio
of $r = 0.2$, where $r$ is defined in terms of the relative amplitudes
of the ensemble averages of the temperature power spectra at $\ell=10$,
\begin{equation}
r = {C^{T_{tensor}}_{\ell = 10} \over C^{T_{scalar}}_{\ell = 10}}.
 \label{TS1} 
\end {equation}
As in E04, unless otherwise stated, beam functions will not be written
explicitly in equations and so $C_\ell$ will sometimes mean $C_\ell
b^2_\ell$, where $b_\ell$ is the Gaussian beam function
\begin{equation}
b_\ell  = {\rm exp} \left ( -{1 \over 2} \ell (\ell+1) (0.425\theta_s)^2 \right ).
 \label{BF1} 
\end {equation}
 
Figure 1 also shows one of the Kpn family of WMAP masks (see Bennett
\etal, 2003b, for a discussion of the WMAP masks). The Kp0 mask shown
in Figure 1 removes about 21 per cent of the sky at low Galactic
latitudes and is a relatively conservative Galactic mask. (For
reference, the WMAP TE analysis reported by Kogut \etals (2003) used
the less conservative Kp2 mask which removes around 13 per cent of the
sky).

Figure 2 shows the averages of PCL estimates $\tilde C_\ell$ and $\hat
C_\ell$ computed from $10^5$ noise-free simulations with the same
cosmological parameters used to construct Figure 1 and with the Kp0
sky mask applied.  For the Kp0 mask, the matrix $M$ in equation
(\ref{PCL8}) is non-singular and can be inverted. Figure 2 shows that
the PCL estimates $\hat C_\ell$ provide  unbiased estimates of
the true power spectra.

\begin{figure*}

\vskip 4.8 truein

\includegraphics{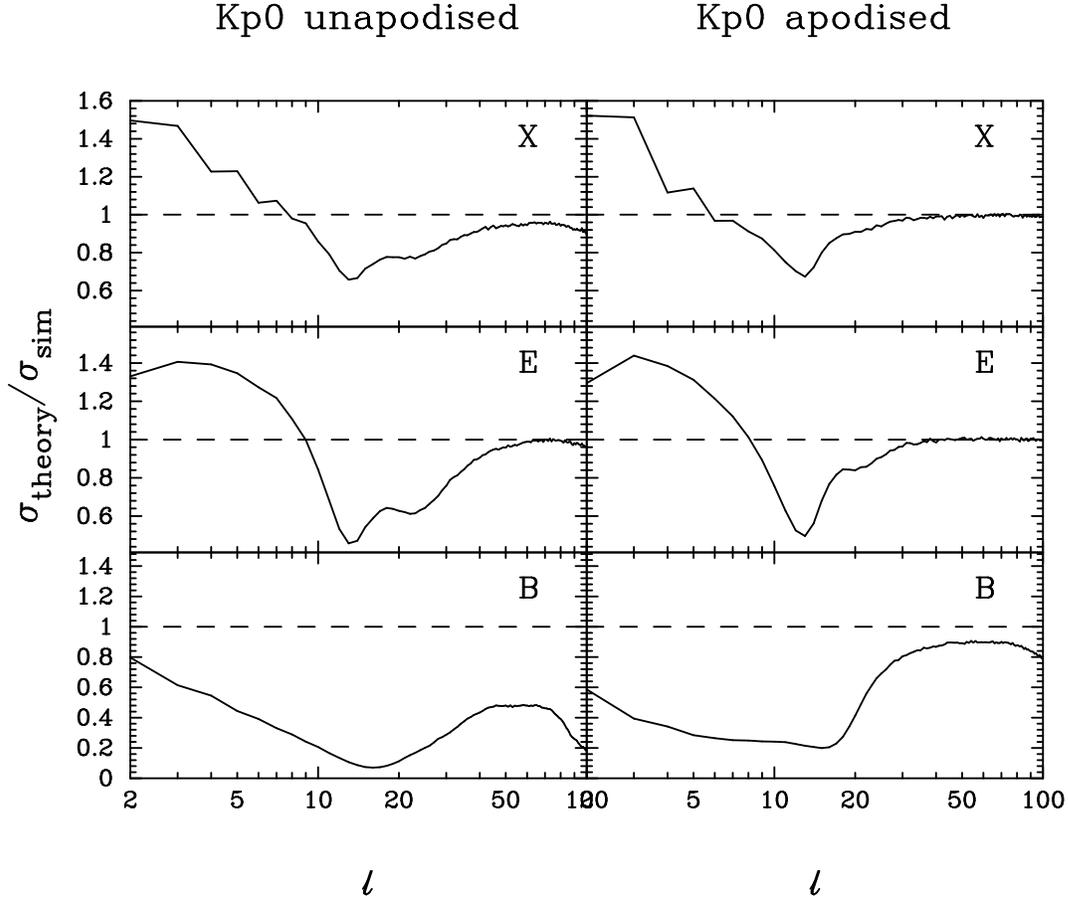}

\caption
{The diagonal components of the PCL power spectrum covariance matrix
estimated from numerical simulations compared to the theoretical
dispersions given in equations (\ref{V1b}) -- (\ref{V1d}).  The
diagrams to the left show results for the unapodised Kp0 mask and
those to the right show results for the apodised Kp0 mask illustrated
in Figure \ref{figure4}. }

\label{figure3}

\end{figure*}

For the temperature polarization power spectrum, accurate expressions
for the covariance matrices of PCL power spectrum at high multipoles
($\ell \gg \Delta L$) can be derived quite easily (see E04). The 
analogous problem for polarization estimates is much more difficult
(see CC05) because the covariances depend on the matrix
products $\;_{\pm}I_{(\ell m)(LM)}\;_{\pm}I_{(LM)(\ell^\prime m^\prime)}$,
$\;_{\pm}I_{(\ell m)(LM)}\;_{\mp}I_{(LM)(\ell^\prime m^\prime)}$, where
\begin{equation}
\;_{\pm}I_{(\ell m)(\ell^\prime m^\prime)} = 
\int {1 \over 2} d{\bf \hat n} w({\bf \hat n})
( \;_{2}Y^*_{\ell m}({\bf \hat n})\;_{2}Y_{\ell^\prime m^\prime}({\bf \hat n}) \pm 
\;_{-2}Y^*_{\ell m}({\bf \hat n})\;_{-2}Y_{\ell^\prime m^\prime}({\bf \hat n})) . \label{CC1}
\end{equation}
CC05 show that these products can be expressed as integrals
$\;_{\pm}I_{(\ell m)(\ell^\prime m^\prime)}$ as in
equation (\ref{CC1}), but with the window finction $w$ replaced by
$w^2$ and  integrals involving the gradients of $w$. For
PCL estimates from noise free data, the covariance matrices can
therefore be approximated by terms which depend on the coupling
matrices in equations (\ref{PCL6a} - \ref{PCL6d}), but with $\tilde
W_\ell$ replaced by the power spectrum of the square of the window
function,
\begin{equation}
 \tilde w^2_{\ell m} = \sum  w^2_i 
\Omega_i Y^*_{\ell m}(\theta_i),    \label{CC2}
\end{equation}
together with complicated terms that depend on gradients of the
window functions $w_i$.\footnote{There are no such gradient terms for
the temperature covariance matrix $\langle \Delta \tilde C^T_\ell
\Delta \tilde C^T_{\ell^\prime} \rangle$, see E04 for a detailed
analysis.} Ignoring the gradient terms, the covariances can be
approximated by 
\beglet
\begin{equation}
 \langle \Delta \tilde C^T_\ell 
\Delta \tilde C^T_{\ell^\prime} \rangle 
\approx {2 C^T_\ell C^T_{\ell^\prime} \over (2 \ell^\prime + 1)}
      M^T_{\ell \ell^\prime}, \label{V1a}
\end{equation} 
\begin{equation}
 \langle \Delta \tilde C^X_\ell 
\Delta \tilde C^X_{\ell^\prime} \rangle 
\approx { (C^T_\ell C^T_{\ell^\prime} C^E_\ell C^E_{\ell^\prime})^{1/2}
 \over (2 \ell^\prime + 1)} M^X_{\ell \ell^\prime} +
{ C^X_\ell C^X_{\ell^\prime} \over (2 \ell^\prime + 1)}
      M^T_{\ell \ell^\prime},  \label{V1b}
\end{equation} 
\begin{equation}
 \langle \Delta \tilde C^E_\ell 
\Delta \tilde C^E_{\ell^\prime} \rangle 
\approx { 2C^E_\ell C^E_{\ell^\prime}
 \over (2 \ell^\prime + 1)} M^{EE}_{\ell \ell^\prime},  \label{V1c}
\end{equation} 
\begin{equation}
 \langle \Delta \tilde C^B_\ell 
\Delta \tilde C^B_{\ell^\prime} \rangle 
\approx { 2C^B_\ell C^B_{\ell^\prime}
 \over (2 \ell^\prime + 1)} M^{BB}_{\ell \ell^\prime},  \label{V1d}
\end{equation} 
\begin{equation}
 \langle \Delta \tilde C^E_\ell 
\Delta \tilde C^B_{\ell^\prime} \rangle 
\approx {  [(C^E_\ell C^E_{\ell^\prime})^{1/2} + 
(C^B_\ell C^B_{\ell^\prime})^{1/2} ]^2 
 \over 2 (2 \ell^\prime + 1)} M^{EB}_{\ell \ell^\prime}.  \label{V1e}
\end{equation} 
\endlet 
The covariance matrix of the PCL estimates $\hat C_\ell$ is given by
\begin{equation}
  \langle   \Delta  \hat C_\ell \Delta \hat  
C_{\ell^\prime} \rangle = M^{-1}
\langle  \Delta \tilde C_\ell \Delta \tilde C_{\ell^\prime} 
\rangle (M^{-1})^T,  \label{PCL9}
\end{equation}
where $M$ is the matrix appearing in equation (\ref{PCL5}) ({\it i.e.}
computed from $\tilde W_\ell$, rather than from the power spectrum of
$w_i^2$).  Neglect of the gradient terms will be referred to somewhat
loosely as the `scalar approximation', since the expressions
(\ref{V1a}-\ref{V1e}) depend on only the spin-0 multipoles of the
square of the window function (\ref{CC2}).  It is interesting to ask
under what circumstances the scalar approximation provides accurate
estimates of the covariances. For relatively small sky cuts, such as
the Kp0 mask, these expressions provide quite accurate estimates for
the $X$ and $E$ power spectra at multipoles $\ell \simgt \Delta L$,
where $\Delta L$ is the mode-coupling scale introduced by the sky cut.
The left-hand panels in Figure \ref{figure3} show the diagonal
components of the covariance matrices for the $\tilde C^X_\ell$,
$\tilde C^E_\ell$ and $\tilde C^B_\ell$ power spectra estimated from
equations (\ref{V1b}) - (\ref{V1d}) compared to estimates from the
$10^5$ simulations used to generate Figure 2. For the Kp0 mask, the
characterstic coupling scale is $\Delta L \sim 20$, and one can see
from Figure 3 that the simple analytic expressions provide quite
accurate estimates of the errors of the $X$ and $E$ power spectra at
multipoles $\ell \simgt 50$, but fail by a factor of $\sim 2$ or more
for the $B$ component. The failure of simple analytic expression for
the $B$-component is related to mixing of the $E$ and $B$ modes. For
the $E$ power spectrum, mixing of $E$ and $B$ modes is always
unimportant at high multipoles, $\ell \gg \Delta L$ (since the
$E$-mode amplitude is fixed by the dominant scalar mode). We
therefore expect the scalar approximation to provide accurate
estimates of the covariances for the $X$ and $E$ power spectra at high
multipoles. However, if the intrinsic amplitude of the $B$ mode is
low, mixing of $E$ and $B$ modes can dominate the estimates of $\tilde
C^B_\ell$ and  the covariances $ \langle \Delta \tilde C^B_\ell
\Delta \tilde C^B_{\ell^\prime} \rangle$ and $\langle \Delta \tilde
C^E_\ell \Delta \tilde C^B_{\ell^\prime} \rangle$ at high
multipoles. For our chosen tensor-scalar amplitude, $r =0.2$,
mode-mixing dominates the $\tilde C^B_\ell$ amplitude at high
multipoles, which is why the scalar approximation fails so badly.

\begin{figure*}

\vskip 3.7 truein

\includegraphics{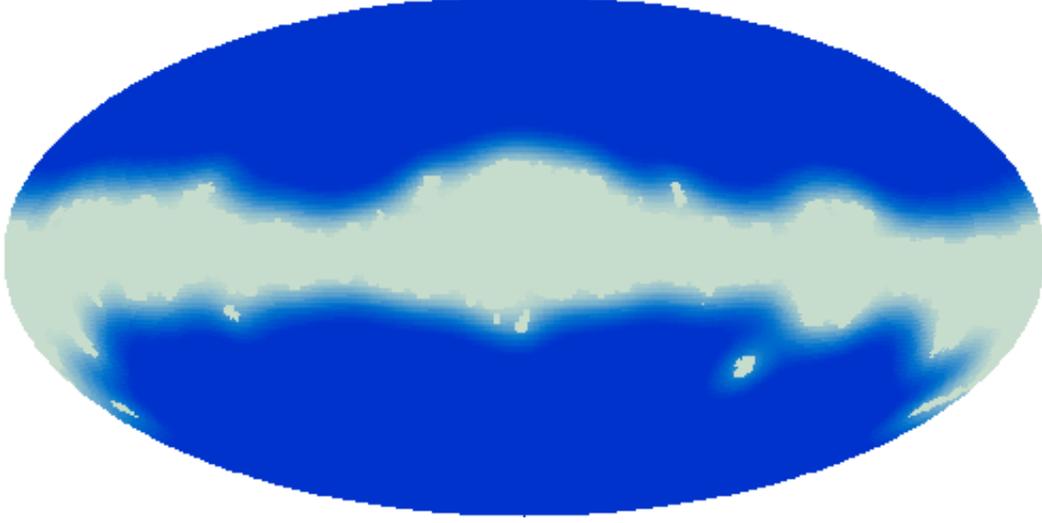}

\caption
{Apodised Kp0 mask at a pixel resolution of $\theta_c = 1^\circ$ created
using the iterative algorithm described in the text.}

\label{figure4}

\end{figure*}

\begin{figure*}
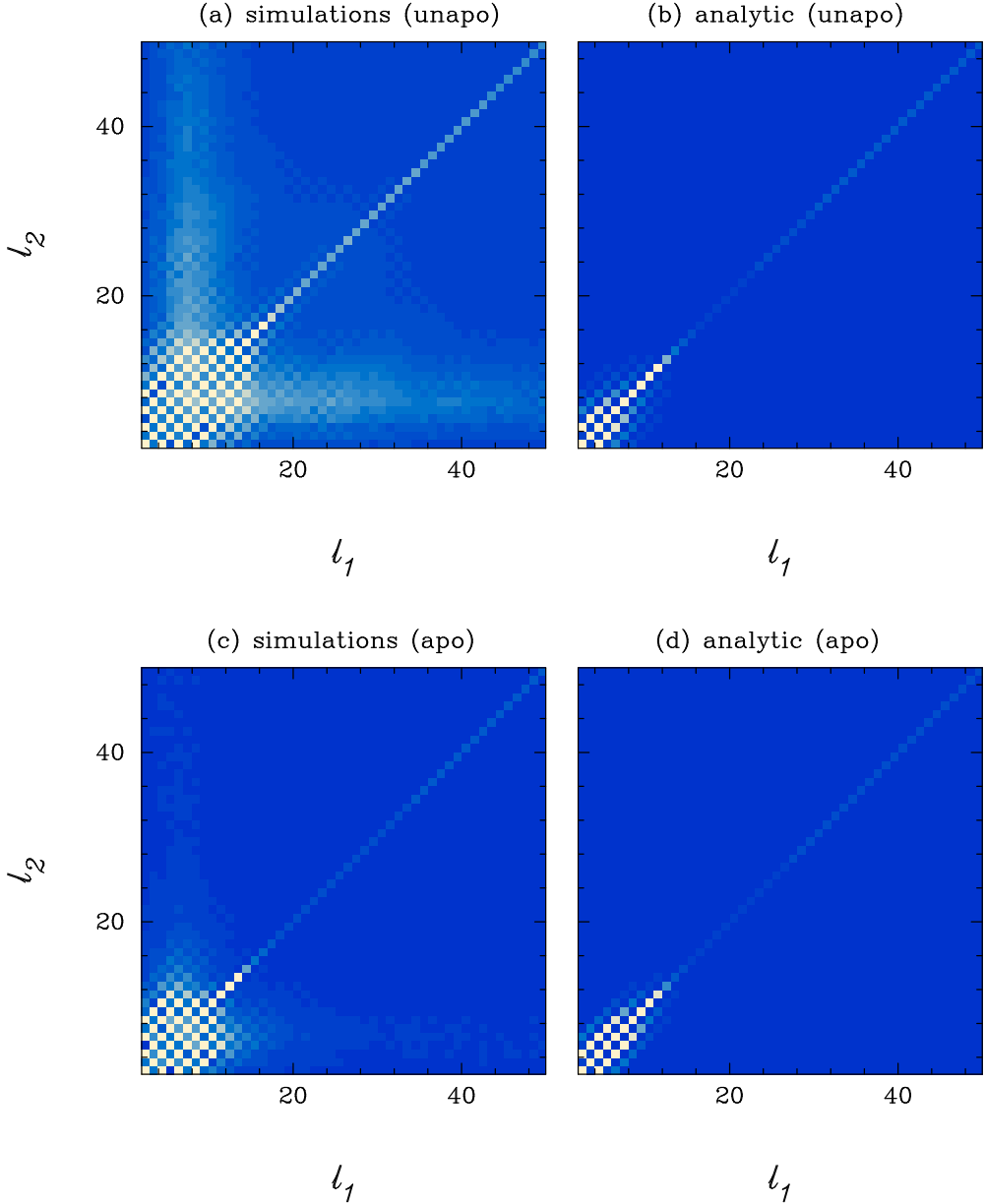


\vskip 6.5 truein

\includegraphics{pgbb_kp0.ps}
\includegraphics{pgbb_kp0w.ps}

\caption
{Covariance matrices for the B-mode polarization spectra, illustrating
the effects of apodization on the PCL estimator. Figures
\ref{figure5}(a) and \ref{figure5}(c) show the covariance matrices
computed from $10^5$ simulations with the same parameters as those in
Figure 1. The simulations used in Figure \ref{figure5}(a) use the
unapodised Kp0 mask while those used in Figure \ref{figure5}(c) use
the apodised Kp0 mask shown in Figure \ref{figure4}. Figures
\ref{figure5}(b) and \ref{figure5}(d) show the corresponding analytic
covariance matrices using the approximation of equation (\ref{V1d})}

\label{figure5}

\end{figure*}

Intuitively, one would expect that it would be possible to
significantly reduce the effects of $E$ and $B$ mode mixing at high
multipoles by apodising the sky cut. Apodisation down-weights data
close to the edge where the non-local nature of the $E$ and $B$ mode
decomposition causes ambiguity between modes ({\it cf} CC05). Thus we
would expect the scalar approximation to be much more accurate if we
apodise the Kp0 mask. Of course, in this context apodisation will lead
to some information loss, but in practice the purpose of applying a
mask is to reduce the effects of systematic errors caused by
inaccuracies in modelling Galactic emission. Hence, the data close to
the edge of a mask will usually be less reliable than data far removed
from the mask. The slight loss of information caused by apodising is
therefore likely to be more than compensated  by the reduction in $E$ and
$B$ mode mixing at high multipoles.

Figure 4 shows an example of an apodisation algorithm applied to the
Kp0 mask. The apodisation algorithm works iteratively as
follows. First, the Kp0 mask in this pixelisation is specified by
$w_i=0$ or $1$ according to whether pixel $i$ lies within or outside
the mask. Then:

\noindent
(i)  compute the spherical transform of the set of
weights, $\tilde w_{\ell m}$, (equation \ref{PCL3e}) and multiply these 
coefficients by a smoothing factor of $b_\ell$ (equation \ref{BF1})
with Gaussian FWHM of $\theta_s = 10^\circ$;

\noindent
(ii) construct a smoothed weight function $\tilde w_i$ by performing an
inverse spherical transform;

\noindent
(iii) set the weights $\tilde w_i$ that lie within the original Kp0
mask  equal to zero;

\noindent
(iv) go back to step (i) and transform the smoothed weights to make a 
new set of coefficients $\tilde w_{\ell m}$ and continue steps (i) -- (iv)
for a specified set of iterations.

The example shown in Figure 4 shows the apodised mask after 4
iterations. The effective smoothing length therefore has a Gaussian
FWHM of $\theta_s = 20^\circ$ and one can see that the `islands'
interior to the masked region evident in Figure 1b have been smoothed
to low amplitudes. The right-hand panels of Figure 3 show the diagonal
components of the covariance matrices for the $X$, $E$ and $B$ mode
PCL power spectra of the apodised maps compared to the predictions of
the scalar approximation. The $X$ and $E$ dispersions now agree almost
perfectly with the scalar approximation for $\ell \simgt 30$. For the
$B$ mode, the variance caused by $E$ and $B$ mode mixing at $\ell
\simgt \Delta L$ is much less than for the unapodised case shown in
Figure 3. For the diagonal components of the $B$ mode power spectrum,
the scalar approximation is accurate to a few percent or so. Figure
\ref{figure5} shows the structure of the $B$ mode covariance matrices
for the unapodised and the apodised Kp0 masks. In the apodised case,
the covariance matrix is diagonally dominant at $\ell \simgt 20$ and
the diagonal components are accurately described by equation
(\ref{V1d}).

 The scalar approximation is therefore a very good approximation for
all three power spectra $X$, $E$ and $B$. Of course, even with
apodisation, the scalar approximation for the $B$ mode spectrum will
fail at some critical value of the tensor-scalar ratio $r$, hence one
cannot simply rescale equation (\ref{V1d}) by the appropriate
amplitude of $C^B_\ell$ for arbitrarily low values of $r$. For low
values of $r$, one can explicitly include the gradient terms which
depend on the amplitude of $C^E_\ell C^E_{\ell^\prime}$ (equation
(80) of CC05) to get an accurate model of the covariance matrix at high
multipoles. However, for most forseable experiments, including {\it
Planck}, instrument noise is likely to dominate the $B$-mode
covariance matrix at high multipoles. In Section 4  we will show that in this 
situation the scalar approximation is extremely accurate independent of the 
intrinsic amplitude of $C^B_\ell$.

%
%
%
%
%
%

\section{Quadratic Maximum Likelihood}

\subsection{Preliminaries}

A QML estimator for temperature CMB power spectra is discussed by
Tegmark (1997). The generalization of this estimator to the estimation of 
CMB polarization is straightforward and is discussed in detail
by TdO01. The estimator will be reviewed here briefly. We work in pixel
space and define an input data vector ${\bf x}$ consisting of the temperature 
differences and Stokes parameters $Q$ and $U$ (defined with respect to
a fixed coordinate system, as in Figure 1) specified at each pixel,
\begin{equation}
{\bf x} \equiv ({\bf \Delta T}, {\bf Q}, {\bf U}).   \label{ML0}
\end{equation} 
The optimal QML power spectrum estimate  is (TdO01)
\begin{equation}
 y^r_{\ell} = x_i x_j E^{r\ell}_{ij}, \qquad r \equiv (T, X, E, B), \label{ML1a}
\end{equation}
where the matrices $E^{r\ell}$ are given by 
\begin{equation}
 E^{r \ell} = 
{1 \over 2}C^{-1} {\partial C \over \partial C^r_\ell} C^{-1},  \label{ML1b}
\end{equation}
and $C$ is the covariance matrix of the data vector ${\bf x}$,
\begin{equation}
 C_{ij}  = \langle x_i x_j \rangle 
 =  {\left ( \begin{array}{ccc}
        C^{TT} & C^{TQ} & C^{TU}  \\
        C^{QT} & C^{QQ} & C^{QU} \\
        C^{UT} & C^{UQ} & C^{UU} 
       \end{array} \right ) }.  \label{ML1}
\end{equation}
The matrices $E^{r \ell}$ defined in equation (\ref{ML1b}) give
formally minimum variance power spectrum estimates if the covariance
matrix $C_{ij}$ is set equal to the true covariance matrix. However,
as noted by TdO01, equation (\ref{ML1a}) mixes $\Delta T$ components
of the data vector ${\bf x}$ with $Q$ and $U$ components in providing
estimates of the $E$ and $B$ mode power spectra. This is undesirable
because for realistic noisy data, systematic errors in the $\Delta T$
measurements could contaminate estimates of the much lower amplitude
$E$ and $B$ mode power spectra.  It is therefore safer to separate the
$\Delta T$ measurements from the $Q$ and $U$ measurements by
`reshaping' the matrix (\ref{ML1}):
\begin{equation}
 \check C_{ij}   
 =  {\left ( \begin{array}{ccc}
        C^{TT} &  0 &  0  \\
        0 & C^{QQ} & C^{QU} \\
        0 & C^{UQ} & C^{UU} 
       \end{array} \right ) }  \label{ML2a}
\end{equation}
and using the matrices 
\begin{equation}
 \check E^{r \ell} = {1 \over 2}\check C^{-1} 
{\partial C \over \partial  C^r_\ell} \check C^{-1}. \label{ML2b}
\end{equation}
in place of the matrices $E^{r \ell}$  in equation (\ref{ML1a}). The estimates
$y^r_\ell$ will give unbiased estimates of the true power spectra $C^s_\ell$,
$s \equiv (T, X, E, B)$,
\beglet
\begin{equation}
 \langle y^r_\ell \rangle 
 =    \check F^{sr}_{\ell \ell^\prime} C^s_{\ell^\prime}, \label{ML3a}
\end{equation}
where
\begin{equation}
 \check F^{sr}_{\ell \ell^\prime} = {1 \over 2}
{\rm Tr} \left [  {\partial C \over \partial  C^s_{\ell^\prime}} \check C^{-1}
{\partial C \over \partial  C^r_{\ell}} \check C^{-1} \right ]. \label{ML3b}
\end{equation}
\endlet By reshaping the covariance matrix, the estimate (\ref{ML1a})
will no longer be minimum variance. However, it is straightforward to
show that for a Kp0-type mask, the increase in variance caused by
using (\ref{ML2a}) instead of (\ref{ML1}) is almost imperceptible.

Following E04 it is useful to define re-scaled power spectra
\begin{equation}
 \tilde C^r_{\ell}   =  y^r_\ell / \sum_{\ell^\prime} \check F^{rr}_{\ell \ell^\prime},  \label{ML4a}
\end{equation}
which have similar shapes to the true power spectra $C^r_\ell$. Furthermore, 
if the matrix $\check F$ is invertible, one can define 
unbiased estimates of the true power spectra via
\begin{equation}
\hat C^r_{\ell} = \check F^{-1} y \label{ML4b}.
\end{equation}
These estimators can be considered the QML analogues to the PCL
estimators defined by equations (\ref{PCL4}) and (\ref{PCL8}).

The covariance matrices of the QML estimates is given by
\begin{equation}
\langle y^r_\ell y^s_{\ell^\prime} \rangle 
- \langle y^r_\ell \rangle \langle y^s_{\ell^\prime} \rangle 
\equiv F^{rs}_{\ell \ell^\prime} =  2 {\rm Tr} \left [ C \check E^{r \ell} 
C \check E^{s \ell^\prime} \right ],  \label{ML5} 
\end{equation}
where $F^{rs}$ is the Fisher matrix. From equation (\ref{ML4b}), the 
covariance matrix of the deconvolved QML estimates is given by
\begin{equation}
 \langle \Delta \hat C_{\ell} \Delta \hat C_{\ell^\prime}
\rangle  = \check F^{-1} F \check F^{-1}.  \label{ML6}
\end{equation}
Notice that if we had used the true pixel covariance matrix (\ref {ML1}), rather
than the `reshaped' form (\ref{ML2a}), the covariance matrix (\ref{ML5}) would 
take the more familiar form
\begin{equation}
 \ F^{sr}_{\ell \ell^\prime} =  {1 \over 2}
{\rm Tr} \left [  {\partial C \over \partial  C^s_{\ell^\prime}}  C^{-1}
{\partial C \over \partial  C^r_{\ell}}  C^{-1} \right ]. \label{ML6a}
\end{equation}

Notice also from equations (\ref{ML1a}) and  (\ref{ML2b}) that the estimator $y_\ell^r$
can be written as 
\begin{equation}
  y_\ell^r = {(2 \ell + 1) \over 2 \Omega^2} \tilde C^{zr}_\ell, \label{MLN1}
\end{equation}
where the $\tilde C^{zr}$ are the PCL power spectra as defined in equation (\ref{PCL4})
but computed from the maps
\begin{equation}
  z_j  = \check C^{-1}_{ji} x_i. \label{MLN2}
\end{equation}
($\Omega$ in equation (\ref{MLN1}) is the solid angle of a single map pixel,
all assumed to be of identical area)
Thus, if one can invert the matrix $\check C^{-1}_{ij}$, the QML
estimator can be computed rapidly by applying fast spherical
transforms to the maps defined by equation (\ref{MLN2}). This is
particularly useful if one wants to compute Monte-Carlo simulations of
the QML estimator to characterise the covariance matrix. For
applications to maps with large numbers of pixels, it may well be more
efficient to use Monte-Carlo simulations to estimate errors, rather
than brute force evaluation of (\ref{ML5}), particularly if the
covariance matrix (\ref{ML6}) is accurately band diagonal (which is
true for Kp0-like sky masks, see Figure 7 below).

\subsection{Covariance Matrix at High Multipoles}

In this Section, we will assume that $C^B_\ell =0$ and derive an
approximation to the covariance matrix of $\hat C^E_\ell$ for high multipoles
and a small sky-cut. The analysis presented here is similar to the analysis
of the QML temperature power spectrum covariance matrix given in Section 3.4
of E04.

If the $B$-mode is absent, all of the information on the polarization anisotropies
is contained in the spin $2$ field $\;_2P_i \equiv (Q_i + iU_i)$. The spin $-2$
field $(Q_i - iU_i)$ contains no additional information. We can therefore construct
a quadratic estimator 
\begin{equation}
 y^E_\ell =  \;_2P_iw_i\;_2P_j^* w_jE_{ij}, \label{ML13} 
\end{equation}
where $E_{ij}$ is as defined in equation (\ref{ML1b}) with,
\begin{equation}
 C_{ij} = \langle  \;_2P_i w_i\;_2P_j^* w_j\rangle  =   \sum_{\ell
 m} C^E_\ell w_i w_j\;_2Y_{\ell m}(i)\;_2Y^*_{\ell m}(j).  \label{ML14} 
\end{equation}
where $w_i$ is an arbitrary pixel weight function. 
If the sky cut is small, it will be a good approximation 
to set the inverse of $C$ to
\begin{equation}
C^{-1}_{ij} \approx \sum_{\ell m}  {\Omega_i \Omega_j \over C_\ell^E} {1 \over w_i w_j}
\;_2Y_{\ell m}(i)\;_2 Y^*_{\ell m}(j)   \label{ML15}
\end{equation}
using the completeness and othogonality relations for the tensorial harmonics.
Hence,
\begin{equation}
C^{-1}_{ij} {\partial C_{jk} \over  \partial C^E_\ell} \approx 
  \sum_m { \Omega_i \over C^E_\ell} {w_k \over w_i}
\;_2Y_{\ell m}(i)\;_2 Y^*_{\ell m}(k)    \label{ML11}
\end{equation}
and the Fisher matrix (\ref{ML3b}) is 
\begin{eqnarray}
F^E_{\ell \ell^\prime} 
&=& {1 \over 2} C^{-1}_{ip} {\partial C_{pk} \over  \partial C^E_\ell}
C^{-1}_{kq} {\partial C_{qi} \over  \partial C^E_{\ell^\prime}} 
 \approx {1 \over 2 C^E_\ell C^E_{\ell^\prime}} 
\sum_{ik}\sum_{m m^\prime} \Omega_i \Omega_k 
\;_2Y_{\ell m}(i) \;_2 Y^*_{\ell m}(k) \;_2Y_{\ell^\prime m^\prime}(k)\;_2 
Y^*_{\ell^\prime m^\prime}(i) \nonumber \\
& = & {(2 \ell + 1)  \over 2 C^E_\ell C^E_{\
ell^\prime}} M^{EE}_{\ell \ell^\prime}  \label{ML12}
\end{eqnarray}
where $M^{EE}$ is the coupling matrix of equation (\ref{PCL6c}).
Equation (34) agrees with equation (\ref{V1c}) and shows that for
small sky-cuts and noise-free data, the QML estimator at high
multipoles ($\ell \gg \Delta L$) is statistically equivalent to the
PCL estimator with equal weight per pixel.

The derivation given here is only valid in so far as the B-mode can be
ignored and the sky cut is small enough (and the multipoles high
enough) that the approximation (\ref{ML15}) is accurate. It is not
straightforward to generalise the analysis presented here to
admixtures of $E$ and $B$ modes.  Given the results of Section 2, we
would expect that at high multipoles and low B-mode amplitudes, the
QML estimator for the B-mode power spectrum will be statistically
equivalent to a PCL estimator with suitably apodised sky masks.  In
other words, the QML estimator will self-consistently find the
appropriate apodisation (via equation \ref{MLN2}) for any given geometry
to minimise $E$ and $B$ mode leakage at each multipole.

\subsection{Covariance Matrix at Low Multipoles}

Generalising  the discussion in Section 3.3 of E04 to polarization, 
we can consider the  vector $x_i$ to consist of the harmonic coefficients
$\tilde a^E_{\ell m}$ and $\tilde a^B_{\ell m}$ measured from noise
free data on the cut sky.
These are related to the true harmonic coefficients $a^E_{\ell m}$ and
$a^B_{\ell m}$ by a coupling matrix
\begin{equation}
 {\left ( \begin{array}{c} 
        \tilde a^E_{\ell m} \\
         \tilde a^B_{\ell m} \end{array} \right ) } = 
 =  {\left  ( \begin{array}{cc}
        -\;_{+}I_{(\ell m)(\ell^\prime m^\prime)} & 
         -i\;_{-}I_{(\ell m)(\ell^\prime m^\prime)}     \\
        -i\;_{-}I_{(\ell m)(\ell^\prime m^\prime)} & \;\;\;_{+}
I_{(\ell m)(\ell^\prime m^\prime)}     \\
       \end{array} \right ) }   {\left ( \begin{array}{c} 
         a^E_{\ell^\prime m^\prime} \\
          a^B_{\ell^\prime m^\prime} \end{array} \right ) },
\label{MLNN1}
\end{equation}
where the integrals $\;_\pm I_{(\ell m)(\ell^\prime m^\prime)}$ are
defined in equation (\ref{CC1}). Evidently, if the coupling matrix in
equation (\ref{MLNN1}) can be inverted, then the true power spectra
can be reconstructed. However, if the power spectra extend to high
values of $\ell$ and $m$, then the coupling matrix will be
singular for a finite sky cut, since it is impossible to reconstruct
modes that lie entirely within the cut. Furthermore, the presence of
boundaries leads to `ambiguous' modes, {\it i.e.} modes which cannot
be assigned as pure $E$ and $B$ modes. As discussed in the
Introduction, various authors have investigated harmonic mode
reconstruction from data on an incomplete sky (Lewis \etals 2002; Bunn
\etals 2003; Lewis 2003). The analysis by Lewis (2003) is particularly
relevant to the discussion because he shows how to construct
projections of pure $E$ and $B$ modes for noise-free data with a Kp2
sky mask. For low values of $\ell$, ($\ell \simlt 1/\theta_{\rm cut}$,
where $\theta_{\rm cut}$ is the characteristic width of the sky cut),
the true $E$ and $B$ modes can be reconstructed accurately.  ({\it
i.e.} for suitably band-limited data the coupling matrix (\ref{MLNN1}) is
invertible. As in the case of temperature anisotropies discussed by
E04, for noise-free data on a cut-sky the QML estimator will
effectively invert the coupling matrix at low multipoles,
reconstructing almost the true values of the $E$ and $B$ mode power
spectra for the full sky. The covariance matrix of the QML estimators
at low multipoles should therefore be given by cosmic variance, {\it
i.e.} diagonal with components
\begin{equation}
\langle (\Delta \hat C^r_\ell)^2 \rangle = 
{2 (C_\ell^r)^2 \over (2 \ell + 1)}.
\label{MLNN2}
\end{equation}
At high multipoles, the variance of the QML estimates will increase
above cosmic variance (as given by equation (\ref{ML12}) for the E-mode)
to
\begin{equation}
\langle (\Delta \hat C^E_\ell)^2 \rangle \approx
{2 (C_\ell^E)^2 \over (2 \ell + 1)f_{\rm sky}},
\label{MLNN3}
\end{equation}
where $f_{\rm sky}$ is the fraction of the sky surveyed, reflecting
the loss of information on modes within the sky cut. (See CC05
for the analogue of equation 
(\ref{MLNN3}) for $\langle( \Delta \hat C^B_\ell)^2
\rangle$ at high multipoles in the limit $C^B_\ell = 0$.) The
expectations of equations (\ref{MLNN2}) and (\ref{MLNN3}) are borne
out by the tests described in Sections 3.4 and 5.

\subsection{Comparison of PCL and QML methods for the Kp0 mask}

Using the `reshaped' covariance matrix of equation (\ref{ML2a}), the
problem of estimating $E$ and $B$ mode power spectra becomes decoupled
from the problem of estimating $T$ and $X$ power spectra. We will
therefore confine the discussion in this Section to the estimation of
$E$ and $B$ mode power spectra using the QML methodology described
in Section 3.1. To compare the QML and PCL methods we have performed a set of
$10^5$ noise-free simulations of the concordance $\Lambda$CDM model
with $5^\circ$ pixels and a smoothing of $\theta_s=7^\circ$. The Kp0
mask was applied, leaving $1216$ active pixels out of a total of $1632$
pixels over the full sky. For these simulations, the amplitude of the
input $B$ mode power spectrum was set to zero. For the PCL estimates, the
Kp0 mask was apodised as described in Section 2 (see Figure \ref{figure4}). 

\begin{figure*}
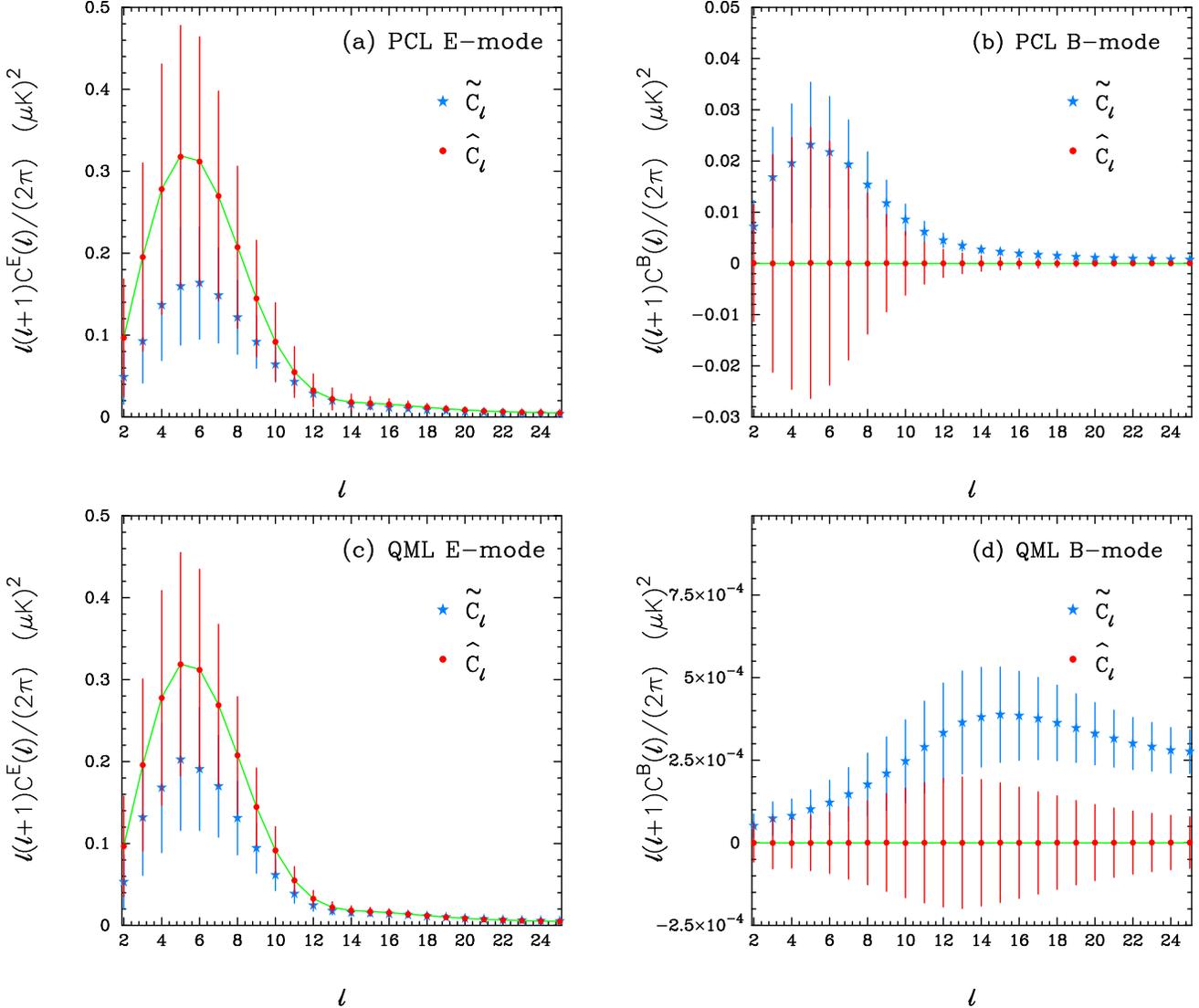


\vskip 6.0 truein

\includegraphics{pgquadeb_a.ps}
\includegraphics{pgquadeb_c.ps}
\includegraphics{pgquadeb_b.ps}
\includegraphics{pgquadeb_d.ps}

\caption
{Comparision of PCL and QML polarization power spectrum estimates
using the simulations described in Section 3.4. These simulations are
noise-free, smoothed with a Gaussian of FWHM $\theta_s = 7^\circ$ and
have had the Kp0 mask applied. In the case of the PCL estimates, an
apodised Kp0 mask was applied using the algorithm described in Section
2 (see Figure \ref{figure4}). The (blue) stars in each panel show the
convolved estimates $\tilde C_\ell$ and the (red) circles show the
deconvolved estimates $\hat C_\ell$. The (green) lines in Figures (a)
and (c) show the $C^E_\ell$ spectrum of the $\Lambda$CDM model.  The
input $C^B_\ell$ spectrum was assumed to be zero. Error bars show
$1\sigma$ errors. Note the large difference in scales on the diagrams
showing the PCL and QML B-mode power spectrum estimates. }

\label{figure6}

\end{figure*}

\begin{figure*}
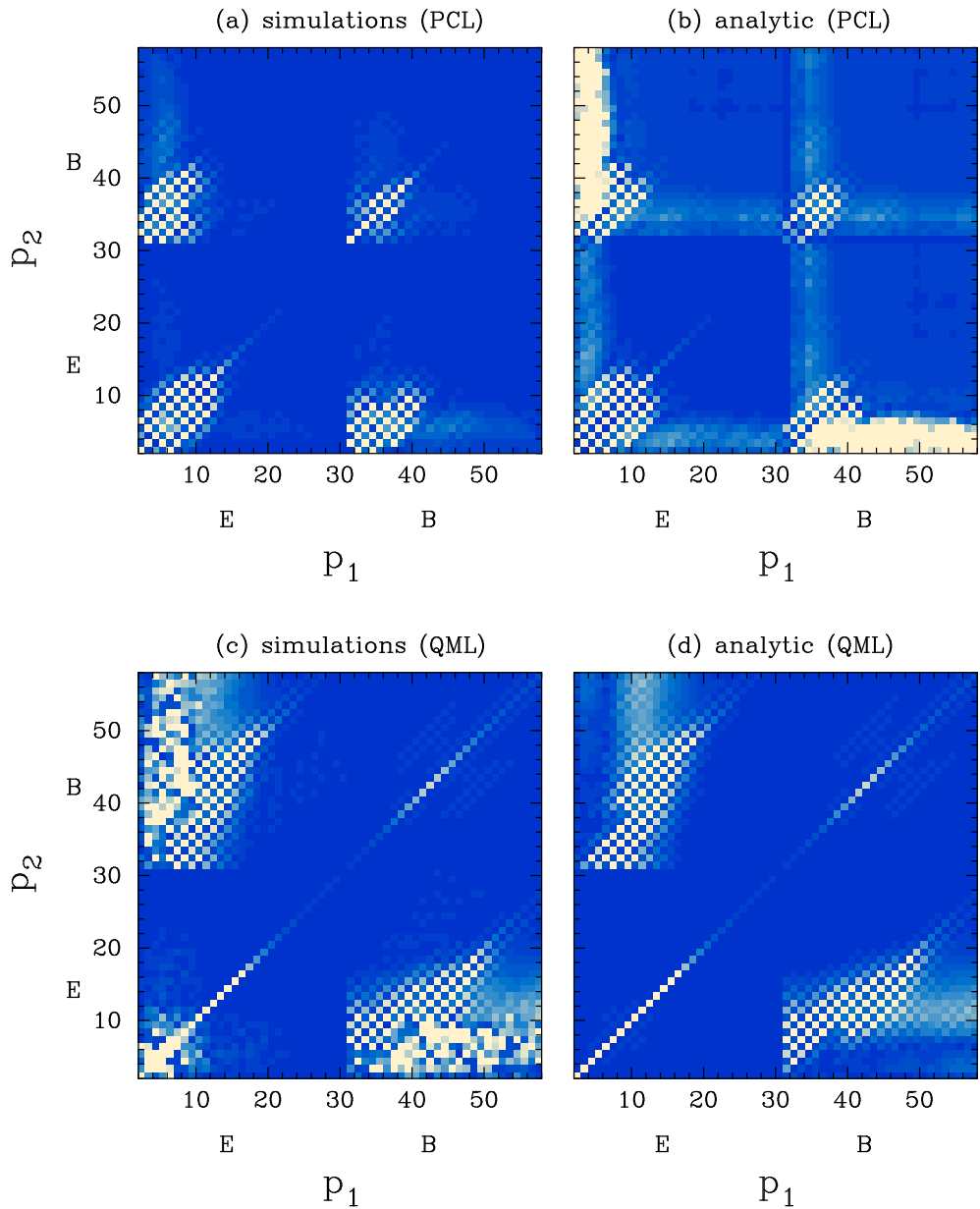


\vskip 6.5 truein

\includegraphics{pgcovquad1.ps}
\includegraphics{pgcovquad2.ps}

\caption
{Comparison of the covariance matrices for PCL and QML estimates
of the E and B mode polarization spectra. The upper diagrams 
show the covariance matrices of the `deconvolved' PCL estimates
$\hat C^E_\ell$ and $\hat C^B_\ell$ (equation \ref{PCL8}) 
derived from numerical simulations (see Figure 7) compared with the
analytic predictions of equations (\ref{V1c})-- (\ref{V1e}). 
The lower diagrams show equivalent plots for the QML
estimates $\hat C_\ell$ (equation \ref{ML4b}) compared
with the analytic estimates derived from equation (\ref{ML6}).
(Note that the EB quadrants of these diagrams have been rescaled by
large factors to render them visible.)}

\label{figure7}

\end{figure*}

Results from these simulations are shown in Figures \ref{figure6}
and \ref{figure7}. Figure {\ref{figure6} shows that the deconvolved
PCL and QML estimates provide unbiased estimates of the input
power spectra. The errors of the $E$-mode spectra are dominated by
cosmic variance, though one can see that the Kp0 sky cut causes a
noticeable increase in the PCL errors. This is equivalent to the
discussion of `estimator induced' variance for the temperature
anisotropies at low multipoles discussed in Section 3.3 of E04 (see
also Efstathiou 2004b). At low multipoles ($\ell \simlt 10$) the QML
estimator returns very nearly the exact value of $C_\ell^E$ for each
realisation and so the covariance matrix is accurately diagonal
(Figures {\ref{figure7}c and {\ref{figure7}d) and limited by cosmic
variance over the full sky (equation \ref{MLNN2}). In contrast, the PCL
estimates of $C^E_\ell$ at low multipoles are strongly correlated
(Figure \ref{figure7}a) as a result of the sky cut and have a large
estimator-induced variance which amplifies the errors above those of
the QML estimator.

The contast between the PCL and QML $B$-mode estimates is even more
extreme. The input $B$-mode spectrum in these simulations is set to
zero, thus any measured $B$-mode is arising from $E$ mode leakage
caused by the sky cut. From Figures \ref{figure6}b one can see that
for the PCL estimator, leakage leads to $B$-mode amplitudes of $\sim
0.01$--$0.02$ $(\mu{\rm K})^2$, {\it i.e.} about $10 \%$ of the
amplitude of the $E$-mode spectrum, simply as a result of the sky cut.
The estimator induced variance is therefore huge and so even if one
had noise-free data, it would not be possible to probe tensor-scalar
ratios of $r \ll 0.1$ using a Kp0 sky cut and PCL power spectrum
estimates (see also CC05 for a similar conclusion).  As expected from
the discussion in Section 3.3, the QML estimator recovers almost the
exact input power spectrum at low multipoles even on the cut sky. Thus
in Figure \ref{figure6}d the QML $B$-mode estimates have an
amplitude of typically $\simlt 10^{-4}$ $(\mu{\rm K})^2$ and the
covariance matrix for these estimates is accurately diagonal (Figure
\ref{figure7}d). The residual `estimator induced' variance is a
consequence of the coupling between low $\ell$ modes and high $\ell$
modes which limits the accuracy of an inversion of
equation (\ref{MLNN1}). Thus, for a Kp0-type sky cut the variance of the QML
estimator leads to a limit on the detectability of a tensor-scalar
ratio of $r \sim 10^{-5}$ in the absence of noise or residual
foreground emission (see also Amarie, Hirata and Seljak 2005).  It
would be possible to probe lower values of $r$ by isolating pure
$B$-modes on the cut sky, as described by Lewis (2003), though methods
for probing $r \simlt 10^{-5}$ are probably only of academic interest
at this stage, given our lack of knowledge of polarised foregrounds.

\section{Including Instrumental Noise}

\subsection{PCL estimator with noise}

In this Section we consider the effects of simple models of instrument noise. The data
vector (\ref{ML0}) is therefore split into a signal contribution $x^s$ and a noise
contribution $x^n$, which are assumed to be uncorrelated with covariance matrices
$S$ and $N$:
\begin{equation}
x_i = x_i^s+ x_i^n,  \qquad  S_{ij} = \langle x^s_i x^s_j \rangle, \qquad N_{ij} = 
\langle x^n_i x^n_j \rangle. \label{N1}
\end{equation}
We assume further that the noise on each of the $T$, $Q$ and $U$ maps is uncorrelated.
With these assumptions, the expectation values of the PCL power spectrum estimates 
is modified from (\ref{PCL5}) to 
\beglet
\begin{equation}
\langle \tilde C^T_\ell \rangle = M^TC^T + N^T_\ell, \label{N2a}
\end{equation}
\begin{equation}
\langle \tilde C^X_\ell \rangle = M^XC^X,  \label{N2b}
\end{equation}
\begin{equation}
\langle \tilde C^E_\ell \rangle = M^{EE}C^E + M^{EB}C^B + N^E_{\ell}, \label{N2c}
\end{equation}
\begin{equation}
\langle \tilde C^B_\ell \rangle = M^{BB}C^B + M^{EB}C^E + N^B_{\ell}, \label{N2d}
\end{equation}
\endlet
where the noise power spectra $N^T_\ell$, $N^E_\ell$ and $N^B_\ell$ are given by
\beglet
\begin{equation}
N^T_\ell = {1 \over (2 \ell + 1)} \sum_{ijm} Y^*_{\ell m}(i) Y_{\ell m} (j) w_i w_j \Omega_i \Omega_j,  \label{N3a}
\end{equation}
\begin{equation}
N^E_\ell = {1 \over  (2 \ell + 1)} {1 \over 4} \sum_{ijm} (N^Q_{ij}
R^{+*}_{\ell m}(i) R^{+}_{\ell m} (j)  + N^U_{i j} R^{-*}_{\ell m} (i)
R^{-}_{\ell m}(j) ) w_i w_j \Omega_i \Omega_j,  \label{N3b}
\end{equation}
\begin{equation}
N^B_\ell = {1 \over  (2 \ell + 1)} {1 \over 4} \sum_{ijm} (N^Q_{ij}
R^{-*}_{\ell m}(i) R^{-}_{\ell m} (j)  + N^U_{i j} R^{+*}_{\ell m} (i)
R^{+}_{\ell m}(j) ) w_i w_j \Omega_i \Omega_j.  \label{N3c}
\end{equation}
\endlet
If the noise covariance matrices are diagonal,
\begin{equation}
N^T_{ij} = (\sigma^T_i)^2 \delta_{ij}, \quad N^Q_{ij} = (\sigma^Q_i)^2 \delta_{ij}, \quad N^U_{i,j}
 = (\sigma^Q_i)^2 \delta_{ij},   \label{NN3}
\end{equation}
\endlet
then the expressions (\ref{N3a})-(\ref{N3c}) simplify to 
\beglet
\begin{equation}
N^T_\ell = {1 \over 4 \pi} \sum_{i} (\sigma_i^T)^2 w_i^2 \Omega_i^2, \label{N4a}
\end{equation}
\begin{equation}
N^E_\ell = N^B_\ell = {1 \over  8 \pi} \sum_{i} ( (\sigma_i^Q)^2 
+ (\sigma_i^U)^2) w_i^2 \Omega_i^2,  \label{N4b}
\end{equation}
\endlet
where the last equation follows from the addition relation for tensorial harmonics,
\begin{equation}
\sum_m \;_{s_1} Y^*_{\ell m} (\theta_1, \phi_1) \;_{s_2}Y_{\ell m} (\theta_2, \phi_2)
= \left ( {2 \ell + 1 \over 4 \pi} \right)^{1/2} \;_{s_2} Y_{\ell -s_1} (\beta, \alpha)
{\rm e}^{-is_2 \gamma}, \label {N4}
\end{equation}
and the angles ($\theta_1, \phi_1$) and ($\theta_2, \phi_2$) are related to the 
Euler angles ($\alpha, \beta, \gamma$) as described in Section 1.4.7 of Varshalovich,
Moskalev and Khersonskii (1988).

Defining the coupling matrices,
\beglet
\begin{equation}
  \Xi_T(\ell_1, \ell_2, \tilde W)  =  
\sum_{\ell_3 }  {(2 \ell_3 + 1) \over 4 \pi}\tilde W _{\ell_3}
{\left ( \begin{array}{ccc}
        \ell_1 & \ell_2 & \ell_3  \\
        0  & 0 & 0
       \end{array} \right )^2}, \label{N5a}
\end{equation}

\begin{equation}
  \Xi_X(\ell_1, \ell_2, \tilde W)  =  
\sum_{\ell_3 }  {(2 \ell_3 + 1) \over 8 \pi}\tilde W _{\ell_3} (1 + (-1)^L)
{\left ( \begin{array}{ccc}
        \ell_1 & \ell_2 & \ell_3  \\
        0  & 0 & 0
       \end{array} \right )} {\left ( \begin{array}{ccc}
        \ell_1 & \ell_2 & \ell_3  \\
        -2  & 2 & 0
       \end{array} \right )}, \label{N5b}
\end{equation}
\begin{equation}
  \Xi_{EE}(\ell_1, \ell_2, \tilde W)  =  
\sum_{\ell_3 }  {(2 \ell_3 + 1) \over 16 \pi}\tilde W _{\ell_3} (1 + (-1)^L)^2
       {\left ( \begin{array}{ccc}
        \ell_1 & \ell_2 & \ell_3  \\
        -2  & 2 & 0
       \end{array} \right )^2}, \label{N5c}
\end{equation}
\begin{equation}
  \Xi_{EB}(\ell_1, \ell_2, \tilde W)  =  
\sum_{\ell_3 }  {(2 \ell_3 + 1) \over 16 \pi}\tilde W _{\ell_3} (1 - (-1)^L)^2
       {\left ( \begin{array}{ccc}
        \ell_1 & \ell_2 & \ell_3  \\
        -2  & 2 & 0
       \end{array} \right )^2}, \label{N5d}
\end{equation}
\endlet
for arbitrary $\tilde W$, the analogues of equations (\ref{V1a}) -
(\ref{V1e}) including uncorrelated instrumental noise  in the case
$(\sigma^Q_i)^2 = (\sigma^U_i)^2$ are:
\beglet
\begin{eqnarray}
 \langle \Delta \tilde C^T_\ell \Delta \tilde C^T_{\ell^\prime}
\rangle &  \approx &
2 C^T_\ell C^T_{\ell^\prime} 
      \Xi_T(\ell, \ell^\prime, \tilde W^2) + 
 2\Xi_T(\ell, \ell^\prime, \tilde W^{TT}) + 4 (C^T_\ell
C^T_{\ell^\prime})^{1/2} \Xi_T(\ell, \ell^\prime, \tilde W^{2T}), 
\label{N6a}
\end{eqnarray}
\begin{eqnarray}
\langle \Delta \tilde C^X_\ell \Delta \tilde C^X_{\ell^\prime}
\rangle & \approx  & 
 (C^T_\ell C^T_{\ell^\prime} C^E_\ell C^E_{\ell^\prime})^{1/2}
\Xi_X(\ell, \ell^\prime, \tilde W^2) + 
 C^X_\ell C^X_{\ell^\prime} \Xi_T(\ell, \ell^\prime, \tilde W^2) + 
   \Xi_X(\ell, \ell^\prime, \tilde W^{TQ}) +    \nonumber  \\
 &  & \qquad 
(C^T_\ell
C^T_{\ell^\prime})^{1/2} \Xi_X(\ell, \ell^\prime, \tilde W^{2Q}) + (C^E_\ell
C^E_{\ell^\prime})^{1/2} \Xi_X(\ell, \ell^\prime, \tilde W^{2T}),
\label{N6b}
\end{eqnarray}
\begin{eqnarray}
\langle \Delta \tilde C^E_\ell \Delta \tilde C^E_{\ell^\prime}
\rangle & \approx &  2C^E_\ell C^E_{\ell^\prime} \Xi_{EE}(\ell, \ell^\prime, \tilde W^2)
+   2 \Xi_{EE}(\ell, \ell^\prime, \tilde W^{QQ}) + 
4(C^E_\ell
C^E_{\ell^\prime})^{1/2} \Xi_{EE}(\ell, \ell^\prime, \tilde W^{2Q}),
\label{N6c}
\end{eqnarray}
\begin{eqnarray}
\langle \Delta \tilde C^B_\ell \Delta \tilde C^B_{\ell^\prime}
\rangle & \approx &  2C^B_\ell C^B_{\ell^\prime} \Xi_{EE}(\ell, \ell^\prime, \tilde W^2) +
  2 \Xi_{EE}(\ell, \ell^\prime, \tilde W^{QQ})  + 4(C^B_\ell
C^B_{\ell^\prime})^{1/2} \Xi_{EE}(\ell, \ell^\prime, \tilde W^{2Q}),
\label{N6d}
\end{eqnarray}
\begin{eqnarray}
\langle \Delta \tilde C^E_\ell \Delta \tilde C^B_{\ell^\prime}
\rangle & \approx & 
[(C^E_\ell C^E_{\ell^\prime})^{1/2} + 
(C^B_\ell C^B_{\ell^\prime})^{1/2} ]^2 
 \Xi_{EB}(\ell, \ell^\prime, \tilde W^{2}) +
2  \Xi_{EB}(\ell, \ell^\prime, \tilde W^{QQ}) +  
4(C^E_\ell
C^E_{\ell^\prime})^{1/2}\Xi_{EB}(\ell, \ell^\prime, \tilde W^{2Q})
\label{N6e}
\end{eqnarray}
\endlet
where the various window functions appearing in equations (\ref{N6a})--(\ref{N6e}) are
defined by
\beglet
\begin{equation}
 \tilde W^{TT}_\ell = {1 \over (2 \ell + 1)} \sum_m {\it Re}(
w^{T}_{\ell m}  w^{*T}_{\ell m}), \qquad
 \tilde W^{TQ}_\ell \equiv \tilde W^{TU} =  {1 \over (2 \ell + 1)} \sum_m {\it Re}(
w^{T}_{\ell m}  w^{*Q}_{\ell m}),  \qquad \label{N8a}
\end{equation}
\begin{equation}
 \tilde W^{2T}_\ell = {1 \over (2 \ell + 1)} \sum_m {\it Re}(
w^{(2)}_{\ell m}  w^{*T}_{\ell m}), \qquad
 \tilde W^{2Q}_\ell \equiv \tilde W^{2U} =  {1 \over (2 \ell + 1)} \sum_m {\it Re}(
w^{(2)}_{\ell m}  w^{*Q}_{\ell m}), \label{N8b}
\end{equation} 
\begin{equation}
 \tilde W^{QQ}_\ell \equiv \tilde W^{QU} \equiv 
\tilde W^{UU} =  {1 \over (2 \ell + 1)} \sum_m \vert w^{Q}_{\ell
m} \vert^2, 
\label{N8c}
\end{equation} 
\endlet
and 
\begin{equation} 
 w^{(2)}_{\ell m} = \sum_i 
  w_i^2 \Omega_i  Y_{\ell m} (i), \quad  w^{T}_{\ell m} = \sum_i 
 (\sigma^T_i)^2 w_i^2 \Omega_i^2  Y_{\ell m} (i),\quad  w^{Q}_{\ell m} 
\equiv w^U_{\ell m} = \sum_i 
 (\sigma^Q_i)^2 w_i^2 \Omega_i^2  Y_{\ell m} (i). \label{N7}
\end{equation}

\begin{figure*}

\vskip 4.9 truein

\includegraphics{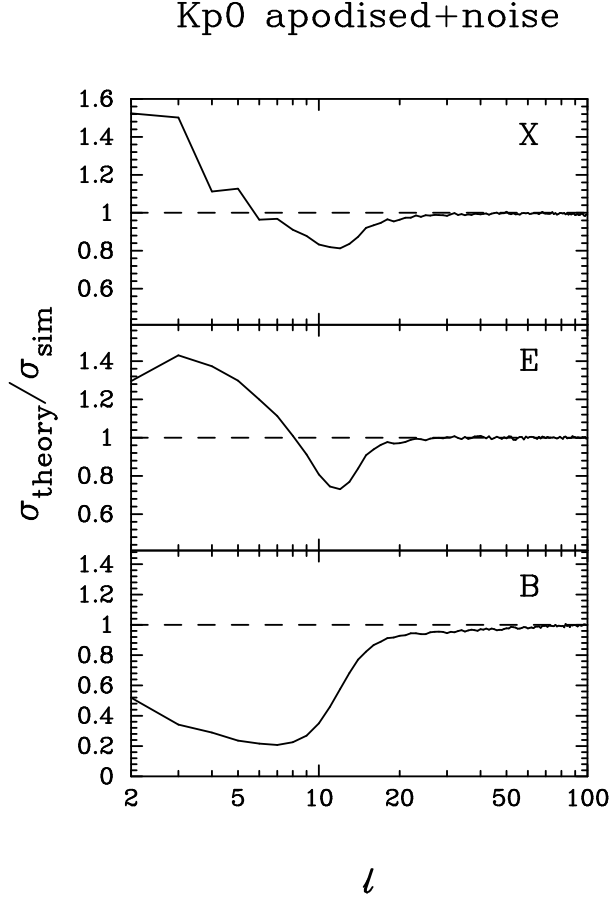}

\caption
{The diagonal components of the PCL power spectrum covariance matrix
estimated from numerical simulations including instrumental noise
(Section 4.1) compared to the theoretical dispersions
given by equations  (\ref{N6b}) -- (\ref{N6d}). Apart from the addition of
noise, the parameters of the simulations were identical to those used
to generate Figure 3.}

\label{figure8}

\end{figure*}

To check these expressions, we repeated the simulations described in
Section 2 ({\it i.e.} $\theta_c = 1^\circ$, $\theta_s = 2^\circ$,
$r=0.2$) but including uncorrelated noise with $\sigma^Q = \sigma^U =
\sigma^T/\sqrt{2} = 2.725 \mu {\rm K}$. With these noise levels, the
$B$-mode power spectrum is noise dominated for $\ell \simgt 20$.  PCL
power spectrum estimates, using the apodised Kp0 mask of Figure 4,
were computed for $10^5$ simulations.

Figure \ref{figure8} compares the diagonal components of the
covariance matrices for the $X$, $E$ and $B$ power spectra compared to
the analytic expressions of equations (\ref{N6b}) --
(\ref{N6d}). Apart from the addition of instrumental noise, these
simulations are identical to those described in Section 2 and so
Figure \ref{figure8} can be compared directly with Figure 3.  Figure
\ref{figure9} compares the full forms of the covariance matrices computed
from the simulations to the analytic predictions of equations
(\ref{N6b}) -- (\ref{N6e}).  Figures \ref{figure8} and \ref{figure9}
show that that the scalar approximation is extremely accurate for all
three spectra $X$, $E$ and $B$ for $\ell \simgt 20$. In the discussion
of Figure 3, we pointed out that in the noise free case, the scalar
approximation would break down for low tensor amplitudes (when the 
PCL B-mode amplitude is dominated by E-B mode mixing rather than 
true B modes). However, if the B-mode power spectrum is noise dominated
at high multipoles (as it is in the simulations described in this Section),
the covariance matrix is fixed by the pure noise term in equation (\ref{N6d})
which is independent of the amplitude of the intrinsic B mode. Thus in the
noise dominated case, the scalar approximation will provide an accurate
estimate of the PCL covariance matrices at high multipoles independent
of the amplitude of the tensor component.

\begin{figure*}
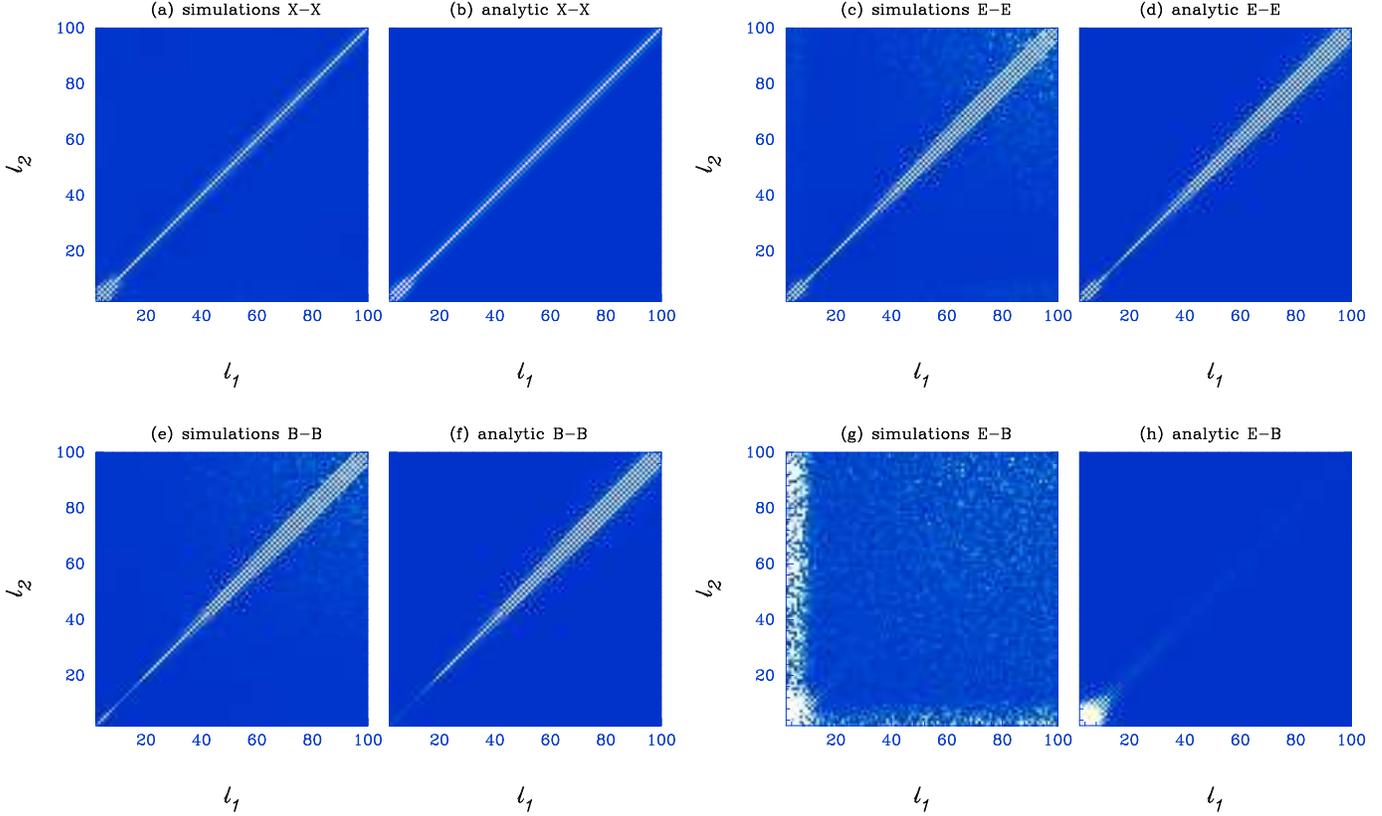


\vskip 4.6 truein

\includegraphics{pgcovxx_kp0wn.ps}
\includegraphics{pgcovbb_kp0wn.ps}
\includegraphics{pgcovee_kp0wn.ps}
\includegraphics{pgcoveb_kp0wn.ps}

\caption
{The PCL $X-X$, $E-E$, $B-B$ and $E-B$ covariance matrices for the simulations 
including instrumental noise (Section 4.1)  compared to the analytic approximations
of equations (\ref{N6b})-(\ref{N6e}).}

\label{figure9}

\end{figure*}

\vskip 1 truein

\subsection{QML estimator with noise}

Evidently, if the data vector $x_i$ includes instrumental noise, the generalization of 
equation (\ref{ML1a}}) is,
\begin{equation}
 y^r_{\ell} = x_i x_j E^{r\ell}_{ij} - N_{ij} E^{r\ell}_{ij}, \label{QMLN1}
\end{equation}
and the covariance matrices $C$ in equations ({\ref{ML1b}) and (\ref{ML1}) are
replaced by $C_{ij} = S_{ij}  + N_{ij}$, where $S$ and $N$ are the signal and
noise covariance matrices defined in equation (\ref{N1}).

If the noise is diagonal, $N^{rs}_{ij} = (\sigma^{r}_i)^2
\delta_{ij}\delta_{rs}$, then in the noise dominated limit
$C^{-1}_{ij} \approx 1/\sigma^2_{i} \delta_{ij}$ and one can see
immediately from equation (\ref{MLN2}) that the QML estimator is
mathematically equivalent to a PCL estimator applied to a map with
inverse variance weighting, $w^r_i = 1/(\sigma^r)^2_i$. As in the
discussion of temperature power spectrum estimation (Hinshaw \etals
2003, E04), we find that the optimal weighting of a PCL estimator
is:

\noindent
(i) equal weight per pixel in the signal dominated limit;

\noindent
(ii) inverse variance weighting in the noise-dominated limit,
if the noise covariance matrix is diagonal.

As described in E04, one of the key motivations for a hybrid estimator
is to construct an estimator that combines PCL power spectra computed
with various weighting schemes to give a near optimal power spectrum
estimate between the two extremes (i) and (ii). This intermediate regime
is difficult to analyse analytically. The hybrid estimator is discussed in the next
Section, which parallels closely the discussion for temperature
anisotropies given in E04. There is, however, an important difference
between the discussion of temperature and polarisation
anisotropies. Consider, for example, the optimal inverse-variance weighted 
$E$-mode pseudo-multipole,
\begin{equation}
\tilde a^E_{\ell m} = 
-{1 \over 2} \sum_i \left ({Q_i \over (\sigma^Q_i)^2} R^{+*}_{\ell m} + 
i{U_i \over (\sigma^U_i)^2} R^{-*}_{\ell m} \right )   \Omega_i
\end{equation}
Because of the spin $\pm2$ nature of the polarisation anisotropies, it
is not possible to derive simple expressions analogous to equation
(\ref{PCL5}) relating PCL power spectra constructed from these
multipoles to the true power spectra unless $(\sigma^Q_i)^2 =
(\sigma^U_i)^2$. If $(\sigma^Q_i)^2 \ne
(\sigma^U_i)^2$, one would need to evaluate
(time-consuming) products of $\ell^2 \times \ell^2$ matrices to recover
the true power spectra. For simplicity, we therefore assume $(\sigma^Q_i)^2 = 
(\sigma^U_i)^2$ in
the rest of this paper, recognising that if this is not true, then it
is not possible to define an easily computable optimal PCL estimator
in the noise-dominated limit.

\section{Hybrid Polarization Power Spectrum Estimator}

Following E04, if we estimate a number of power spectra
$\hat C^{r\alpha}_\ell$ derived from the same data, where $r$ denotes the mode 
($r \equiv T, X, E, B$) and $\alpha$ denotes the estimator ({\it e.g.}
PCL estimates with different weight functions), then they can be combined
into a  single data vector $\hat C^r_{\alpha \ell}$ 
from which we can define a $\chi^2$, 
\begin{equation}
\chi^2 =    (\hat C^r_{\alpha\ell_1} - 
\hat C^{rh}_{\ell_1}){\cal F}^{rs}_{\alpha \ell_1 \beta \ell_2}  
(\hat C^s_{\beta \ell_2} - \hat C^{sh}_{\ell_2}),    \label{H3}
\end{equation}
where $\hat C^{rh}_\ell$ is the hybrid estimator for mode $r$, ${\cal
F}^{rs}_{\alpha \ell_1 \beta \ell_2}$ is the inverse of the covariance
matrix $\langle \Delta \hat C^{r}_{\alpha \ell_1} \Delta \hat
C^s_{\beta \ell_2} \rangle$.  Minimising equation (\ref{H3}) gives the
following linear equations, which we can solve to form the hybrid
estimate $\hat C^{rh}_{\ell}$
\begin{equation}
  \sum_{ s \alpha \beta \ell_2} {\cal F}^{rs}_{\alpha \ell_1 \beta \ell_2} \hat
  C^{sh}_{\ell_1} =
\sum_{s \alpha \beta \ell_2} {\cal F}^{rs}_{\alpha \ell_1 \beta \ell} \hat C^s_{\beta \ell_2},    \label{H4}
\end{equation}
with  covariance matrix
\begin{equation}
\langle  \Delta \hat C^{rh}_{\ell_1} \Delta \hat C^{sh}_{\ell_2} \rangle = 
\left (\sum_{\alpha \beta} {\cal F}^{rs}_{\alpha \ell_1 \beta \ell_2} 
\right )^{-1}.
\label{H5}
\end{equation}
Evidently, since the hybrid estimator involves linear combinations
of the power spectrum estimates $\hat C^r_\ell$, it will  provide an
unbiased estimate of the true power spectra, provided that each of the
$\hat C^r_\ell$ is unbiased. Nevertheless,  as with the discussion of the
optimal QML estimator in Section 3.1, the solution of (\ref{H4}) 
mixes power spectrum estimates for different modes,  which may be undesirable
in practice.  For example, for
PCL estimates, the covariance matrix $\langle \Delta \tilde C^T_\ell
\Delta \tilde C^X_{\ell^\prime} \rangle$ is zero by symmetry, thus the
PCL estimates $\hat C^T_\ell$ and $\hat C^X_\ell$ are strictly
independent. However, the covariance matrices  $\langle \Delta \tilde C^T_\ell
\Delta \tilde C^E_{\ell^\prime} \rangle$ and $\langle \Delta \tilde C^T_\ell
\Delta \tilde C^B_{\ell^\prime} \rangle$ are non-zero
and are given by
\beglet
\begin{equation}
 \langle \Delta \tilde C^T_\ell 
\Delta \tilde C^E_{\ell^\prime} \rangle 
\approx {2 C^X_\ell C^X_{\ell^\prime} \over (2 \ell + 1) (2 \ell^\prime + 1)}
      \sum_{m m^\prime} \left \vert \sum_{\ell_1 m_1} K_{(\ell m)( \ell_1 m_1)}
\;_+I^*_{(\ell^\prime m^\prime)( \ell_1 m_1)} \right \vert^2 , \label{NV1a}
\end{equation} 
\begin{equation}
 \langle \Delta \tilde C^T_\ell 
\Delta \tilde C^B_{\ell^\prime} \rangle 
\approx {2 C^X_\ell C^X_{\ell^\prime} \over (2 \ell + 1) (2 \ell^\prime + 1)}
      \sum_{m m^\prime} \left \vert \sum_{\ell_1 m_1} K_{(\ell m) (\ell_1 m_1)}
\;_-I^*_{(\ell^\prime m^\prime)( \ell_1 m_1)} \right \vert^2 , \label{NV2a}
\end{equation} 
where $\;_\pm I^*_{(\ell m)( \ell^\prime m^\prime)}$ are the coupling 
integrals defined in equation (\ref{CC1}) (with $w$ set to
the weight function applied to the $Q$ and $U$ maps) and 
$K_{(\ell m )(\ell^\prime m^\prime)}$ is the coupling integral
\begin{equation}
K_{(\ell m)(\ell^\prime m^\prime)} = 
\int d{\bf \hat n} w({\bf \hat n})
Y^*_{\ell m}({\bf \hat n})Y_{\ell^\prime m^\prime}({\bf \hat n}) , \label{NV3a}
\end{equation} 
\endlet
with $w$ set to the weight function applied to the temperature map. The 
calculation of the full solution of equation (\ref{H4}) is therefore complicated,
and unless the errors are very well understood, there is a danger of introducing
biases by mixing modes with very different amplitudes. Rather than use the
strict minimum variance solution (\ref{H4}), it is preferable to ignore cross
covariances and  to compute for each mode separately the solution of
\begin{equation}
  \sum_{  \alpha \beta \ell_2} {\cal F}^{rr}_{\alpha \ell_1 \beta \ell_2} \hat
  C^{rh}_{\ell_2} =
\sum_{ \alpha \beta \ell_2} {\cal F}^{rr}_{\alpha \ell_1 \beta \ell_2} 
\hat C^r_{\alpha \ell_2},    \label{H6}
\end{equation}
{\it i.e.}
\begin{equation}
\hat   C^{rh}_{\ell} =   \left ( \sum_{\alpha \beta} 
{\cal F}^{rr}_{\alpha \ell \beta \ell^\prime} \right )^{-1}
\left( \sum_{\gamma\delta} {\cal F}^{rr}_{\gamma \ell^{\prime\prime} \delta \ell^\prime} \hat C^r_{\gamma \ell^
{\prime\prime}} \right ) = B^r_{\ell \gamma \ell^{\prime\prime}} \hat C^r_{\gamma \ell^{\prime\prime}}.    \label{H7}
\end{equation}
The estimates (\ref{H7}) will be technically sub-optimal, but for 
practical purposes they 
will be indistinguishably close to optimal. The covariances and cross-covariances of the hybrid estimates
are straightforward to compute and are given by
\begin{equation}
\langle  \Delta \hat C^{rh}_{\ell_1} \Delta \hat C^{sh}_{\ell_2} \rangle = 
B^r_{\ell_1 \alpha \ell} \langle  \Delta \hat C^{r}_{\alpha \ell} \Delta \hat C^{s}_{\beta \ell^\prime} 
\rangle B^{sT}_{\ell_2 \beta \ell^\prime}.
\label{H8}
\end{equation}

\begin{figure*}

\vskip 3.8 truein

\includegraphics{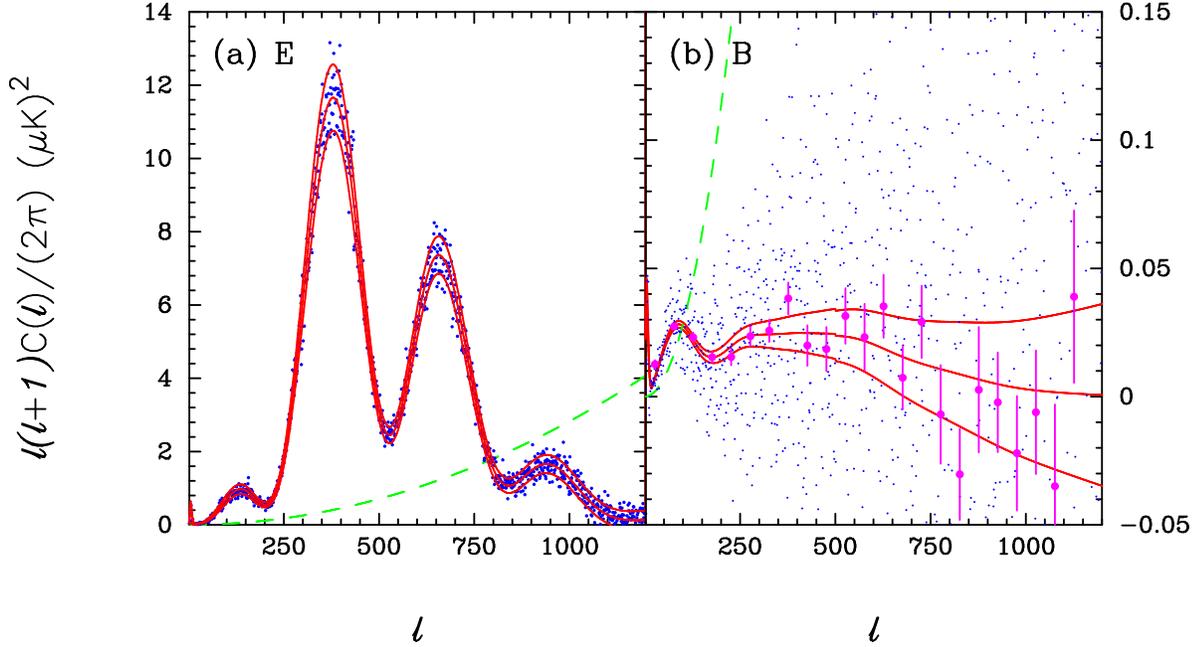}

\caption
{The points show the hybrid estimator applied to PCL estimate of the
$E$ and $B$ mode power spectra as described in the text. The dashed
lines show the white noise level. The lines show the power spectrum of
the fiducial $\Lambda$CDM model and the error ranges given by the
diagonal components of the covariance matrix (\ref{H8}). As the $B$
mode power spectrum is noise dominated for most of the multipole
range, the points with error bars in Figure \ref{figure10}b show the
averages and dispersions of the power spectrum estimates in bands of
width $\Delta \ell = 50$. The theoretical error ranges in this figure
have been scaled appropriately for bands of width $\Delta \ell = 50$.
}

\label{figure10}

\end{figure*}

To illustrate the method, we generated a simulated sky of the
concordance $\Lambda$CDM cosmology with $r=0.2$, at a pixel size of
$\theta_c = 0.1^\circ$ and with a Gaussian smoothing of $\theta_s =
0.25^\circ$. We simulated the hit-count distribution for a Planck-type
scanning strategy as described in E04, {\it i.e.} a single detector
pointing at $85^\circ$ to the spin axis of the spacecraft, and with
the spin axis precessing slowly about the ecliptic plane according to
$5^\circ{\rm sin}(2 \phi_e)$, where $\phi_e$ is the ecliptic
longitude. With this type of scanning strategy, regions with high hit
counts are concentrated at the ecliptic poles as shown in Figure 11 of
E04. The hit count distribution is normalised so that the mean hit
count, $\langle N_{\rm obs} \rangle$ is unity. Uncorrelated Gaussian
noise was then added to the $T$ $Q$ and $U$ maps with $\sigma^Q =
\sigma^U = \sigma^T/\sqrt 2 = 2.725/\sqrt{(N_{\rm obs})} \; \mu{\rm
K}$. 

 We concentrate on the analysis of the $E$ and $B$ power spectra in
this Section.  With these parameters, the $E$ and $B$ mode power spectra
become noise dominated at at $\ell \simgt 750$ and $\ell \simgt 50$ respectively
(see Figure  \ref{figure10}). We apply an apodised Kp0 mask, as shown in Figure 4
but recomputed for the smaller pixels appropriate to these maps. In addition, 
a further weight factor was applied 
\begin{equation}
w_i = {\sigma^2_p \langle \Omega_i \rangle
\over (\sigma^2_i + \epsilon_f \sigma^2_p) \Omega_i}, \label{NE2}
\end{equation}
as described in EO4, where $\sigma^2_p$ is the noise level for a pixel in 
the $Q$ or $U$ map with the mean hit count, and $\sigma^2_i$ is the actual
noise level in pixel $i$. The parameter $\epsilon_f$ controls the weighting
scheme so that $\epsilon_f = 0$ corresponds to inverse variance weighting
and $\epsilon_f \rightarrow \infty$ corresponds to equal weight per pixel
(apart from minor variations in the pixel areas $\Omega_i$ associated
with the pixelisation scheme). 

The results plotted in Figure \ref{figure10} show equation (\ref{H7})
applied to two sets of PCL estimates, one with equal weight per pixel
and one with $\epsilon_f=0.2$. (The cross-covariances between these
estimates can be computed via a straigtforward generalisation of
equations (45) -- (47).)  The solid lines in Figure \ref{figure10}
show the theoretical dispersion computed from the diagonal components of
the covariance matrix (\ref{H8}). As can be seen from the scatter of
the points, these estimates are entirely reasonable. In fact, in this
example, the variances of the PCL estimates at high multipoles are
relatively insensitive to the weighting scheme. At high multipoles,
the $\epsilon_f=0.2$ weighting produces slightly ($\sim 10\%$-$20\%$)
smaller errors compared to equal weight per pixel. However, in the
signal dominated regime at low multipoles, the $\epsilon_f = 0.2$
weighting gives much larger errors. This behaviour is similar to the
behaviour for the temperature anisotropies illustrated in Figure 14 of
E04.

\begin{figure*}
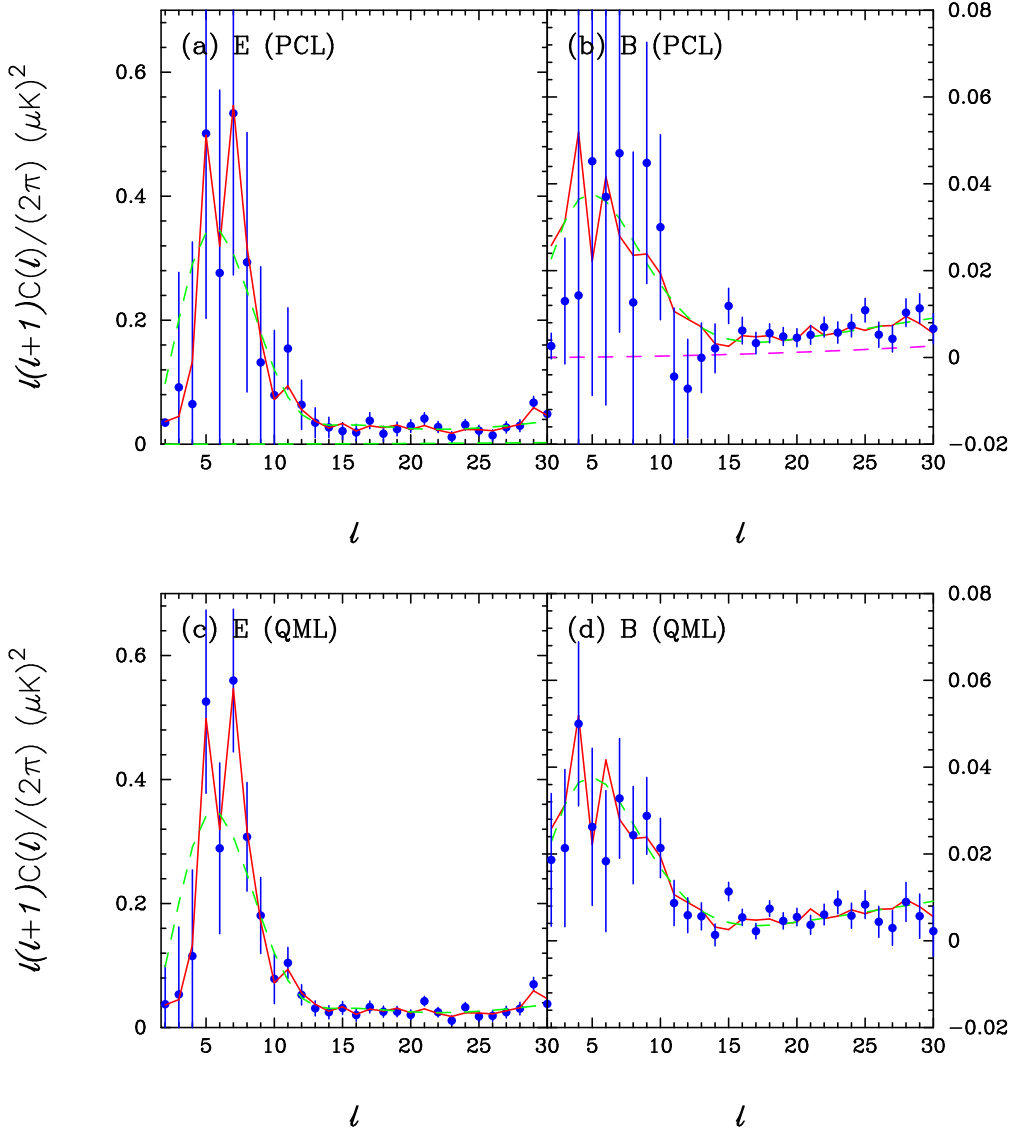


\vskip 6.0 truein

\includegraphics{pg1200smalla.ps}
\includegraphics{pg1200smallb.ps}

\caption
{Figures (a) and (b) show the PCL estimates from Figure \ref{figure10}
at low multipoles on a greatly expanded scale. The error bars are
computed from the diagonal components of the covariance matrix
(\ref{H8}). Figures (c) and (d) show the QML estimates corrected for
the additional smoothing applied to the low resolution maps from which
they were computed.  The solid (red) lines show the input $E$ and $B$
power spectra for this particular simulation, while the dashed (green)
lines shows the power spectra of the fiducial $\Lambda$CDM model. The dashed 
(purple) line in Figure \ref{figure11}(b) shows the noise contribution to
the $B$-mode spectrum. The noise contribution to the $E$ mode spectrum
is negligible at these multipoles.}

\label{figure11}

\end{figure*}

To apply the QML estimator, the maps at high resolution were smoothed
with a Gaussian beam of width $\theta_s = 5^\circ$ and repixelised
onto a low resolution map with pixel size $\theta_c = 3.5^\circ$.  The
QML estimator (\ref{QMLN1}) requires the noise covariance matrix
$N_{ij}$ of the low resolution data. Since the noise in each high
resolution map is diagonal, $\sigma^2_i \delta_{ij}$, the noise
covariance matrix of a low resolution map is given by
\begin{equation}
N_{ij} = \sum_{p \ell \ell^\prime} \sigma^2_p 
(2 \ell + 1)(2\ell^\prime+1) P_\ell({\rm cos} 
\theta_{ip}) P_{\ell^\prime}({\rm cos} \theta_{jp})
{\rm exp} \left( - {1 \over 2} \ell^2 \theta^2_L \right)
{\rm exp} \left( - {1 \over 2} {\ell^\prime}^2 \theta^2_L \right) \Delta \Omega_p^2,
    \label{QMLN2} 
\end{equation}
where $p$ denotes the pixel number in the high resolution map,
$\Omega_p$ is the solid angle of pixel $p$ and $\theta_L$ is the
Gaussian smoothing scale of the low resolution maps. In the small
angle approximation, equation (\ref{QMLN2}) simplifies to the easily
computable expression
\begin{equation}
 N_{ij}   \approx {1 \over 4 \pi^2} {\Delta \Omega_p^2 \over \theta_L^2} \sum_p 
\sigma^2_p {\rm exp} \left( - {\theta_{ip}^2 \over 2 \theta_L^2} \right)
{\rm exp} \left( - {\theta_{jp}^2 \over 2 \theta_L^2} \right), \label{QMLN3}
\end{equation}
which is sufficiently accurate for the example discussed here.

Figures \ref{figure11}a and \ref{figure11}b show the PCL estimates at
low multipoles from Figure \ref{figure10} on a greatly expanded
scale. The error bars on the points show the dispersions computed from
the diagonal components of the covariance matrix The solid (red) lines
show the $E$ and $B$-mode power spectra for this particular simulation
and the  (green) dashed lines show the theoretical power spectra of
the concordance $\Lambda$CDM model. The (purple) dashed line in Figure
\ref{figure11}b shows the theoretical power spectrum of the noise.

Figures \ref{figure11}c and \ref{figure11}d show analogous plots for
 the QML estimator, corrected for the low resolution smoothing
 $\theta_L$. In agreement with the tests shown in Figures
 \ref{figure6} the errors on the QML estimates at low multipoles are
 considerably smaller than those for the PCL estimates. The covariance
 matrices for the QML estimates are very close to diagonal ({\it c.f.}
 Figure \ref{figure7}). Furthermore, the agreement between the $E$ and
 $B$ mode QML power spectrum estimates and the power spectra for this
 particular realisation (shown by the solid red lines), confirms the
 arguments given in Section 3.3 that the QML estimator will return
 almost the exact input power spectra even in the presence of a
 substantial sky cut.  This is similar to the case of temperature
 power spectrum estimation, where a Kp0-like sky cut can introduce a
 substantial `estimator induced' variance (Efstathiou 2004b) at low
 multipoles comparable to the cosmic variance ({\it cf.} Table 1 of
 Efstathiou 2004b). For polarization, however, the problem of
 estimator induced variance is particularly acute because of $E$ and
 $B$-mode mixing. If the $B$-mode amplitude is low, then the estimator
 induced variance of PCL estimators can easily overwhelm cosmic
 variance, whereas it is essentially negligible for a QML
 estimator. Notice that at multipoles $\ell \sim 15$-- $ 25$ the PCL
 and QML estimates are closely similar (as expected from Section
 3.2). At multipoles $\ell \simgt 25$ the errors on the QML estimates
 become larger than those of the PCL estimator because of the large
 smoothing applied to the low resolution maps. However, there is quite
 a wide range of multipoles over which the estimates and errors are
 almost identical. The pixel size and smoothing scale of the low
 resolution simulations used in this example was chosen for
 computational speed so that the QML code could be run quickly on a
 single processor workstation. It would be entirely feasible to extend
 the QML calculation to multipoles of a few hundred or more using
 modern multi-processor computers. For many realistic experimental
 configurations this would provide a large range of overlap with PCL
 estimates at high multipoles. Thus, in practice, it would be
 extremely accurate simply to combine PCL estimates using equation
 (\ref{H7}) and to overwrite the estimates at low multipoles, and
 associated components of the covariance matrices (\ref{H8}), with the
 corresponding QML estimates and covariance matrices.

Alternatively, one can calculate the cross-covariances between 
the QML estimates and PCL estimates, and include the QML estimates
as part of the data vector $\hat C^r_{\alpha \ell}$ used to compute
the hybrid solution $\hat C^{rh}_\ell$. One then requires estimates
of the cross-correlations between the PCL and QML estimates. These
are quite complicated, but for completeness, we give analytic
expressions here for the case of noise-free data.
If we write the QML estimates in the form 
\begin{equation}
y^E_\ell = {\left ( \begin{array}{cc}
        QQ & QU   \\
        UQ  & UU  \\
       \end{array} \right ) } E^{E\ell}  \label{CH6}
\end{equation}
and define the quantities,
\beglet
\begin{equation}
 u^Q_{p\ell m} =  - {1 \over 2} \sum_i (C^Q(\theta_{ip}) R^{+*}_{\ell m} + 
iC^{QU}(\theta_{ip}) R^{-*}_{\ell m}) w_i \Omega_i,    \label{CH7a}
\end{equation}
\begin{equation}
 u^U_{p\ell m} =  - {1 \over 2}  \sum_i (C^U(\theta_{ip}) R^{+*}_{\ell m} + 
iC^{QU}(\theta_{ip}) R^{-*}_{\ell m}) w_i \Omega_i,    \label{CH7b}
\end{equation}
\begin{equation}
 v^Q_{p\ell m} =  - { i \over 2} \sum_i (C^Q(\theta_{ip}) R^{-*}_{\ell m} -
iC^{QU}(\theta_{ip}) R^{+*}_{\ell m}) w_i \Omega_i,    \label{CH7c}
\end{equation}
\begin{equation}
 v^U_{p\ell m} =  - {i \over 2} \sum_i (C^U(\theta_{ip}) R^{-*}_{\ell m} -
iC^{QU}(\theta_{ip}) R^{+*}_{\ell m}) w_i \Omega_i,    \label{CH7d}
\end{equation}
\endlet
then the cross-covariances are given by
\beglet
\begin{equation}
\langle  \Delta \tilde C^E_\ell \Delta  y^r_{\ell^\prime} \rangle = 
{1 \over (2 \ell + 1)} {\left ( \begin{array}{cc}
        \sum_{m} 2u^Q_{p\ell m} u^{Q*}_{q\ell m} &  \sum_m (u^Q_{p\ell m} v^{U*}_{q\ell m}
+ u^{Q*}_{p\ell m}v^{U}_{q \ell m})  \\
 \sum_m (v^U_{p\ell m} u^{Q*}_{q\ell m}
+ v^{U*}_{p\ell m}u^{Q}_{q \ell m})         & \sum_{m} 2v^U_{p \ell m} v^{U*}_{q \ell m}  \\
       \end{array} \right ) } E^{r\ell^\prime}_{pq},  \label{CH8}
\end{equation}
\begin{equation}
\langle \Delta \tilde C^B_\ell \Delta y^r_{\ell^\prime} \rangle = {1
\over (2 \ell + 1)} {\left ( \begin{array}{cc} \sum_{m} 2v^Q_{p\ell m}
v^{Q*}_{q\ell m} & \sum_m (v^{Q}_{p\ell m} u^{U*}_{q\ell m} +
v^{Q*}_{p\ell m}u^{U}_{q \ell m}) \\ \sum_m (u^{U}_{p\ell m}
v^{Q*}_{q\ell m} + u^{U*}_{p\ell m}v^{Q}_{q \ell m}) & \sum_{m}
2u^U_{p \ell m} u^{U*}_{q \ell m} \\ \end{array} \right ) }
E^{r\ell^\prime}_{pq}.  \label{CH9}
\end{equation}
\endlet 
As explained above, for most purposes, it should be
sufficiently accurate simply to overwrite PCL estimates and
covariance matrices at low multipoles with those computed
from QML, avoiding the complexity of evaluating equations 
(\ref{CH8}) and (\ref{CH9}).

\section{Conclusions}

This paper generalises the analysis of power spectrum estimators
presented in E04 to the case of polarisation. Simple analytic
expressions involving the power spectrum of the square of the window
function (the scalar approximation) are given for the covariance
matrices of PCL estimates, including the effects of uncorrelated
instrument noise. For noise free data, the scalar approximation is
shown to give accurate error estimates for the $E$-mode power spectrum
at high multipoles for realistic (Kp0-like) sky-cuts, and also for the
$B$-mode power spectrum provided the Q and U maps are apodised
appropriately and the amplitude of the $B$-mode is high enough. For
noise dominated data, the scalar approximation povides extremely
accurate covariance estimates for both the $E$ and $B$ modes at high
multipoles.

The results presented in Section 3 establish certain relationships
between QML and PCL estimators, in particular, the statistical
equivalence of QML and PCL estimators in the noise-free and
noise-dominated limits.  This analysis parallels the discussion of
temperature power spectrum estimates given in E04. However, in the
noise-dominated limit, an optimal PCL polarization estimator can only be
constructed for the special case of uncorrelated noise and identical
noise levels in the Q and U maps, $(\sigma^Q)^2_i =
(\sigma^U)^2_i$. Fortunately, for a {\it Planck}-type experiment, both
of these assumptions should be reasonably accurate at high multipoles.

An analytic discussion of the noise properties of CMB temperature maps
for a {\it Planck}-like scanning strategy with realistic `$1/f$' noise
is given by Efstathiou (2005). For a {\it Planck}-type experiment, the
noise should be accurately white at high multipoles (see also Stompor
and White, 2004). Low frequency `$1/f$' noise introduces striping in
the maps with a characterstic spectrum that varies approximately as
$\Delta C_\ell \propto 1/\ell$ at low multipoles. This low frequency
behaviour for the temperature power spectra has been verified in many
simulations of {\it Planck}-like experiments ({\it e.g.} Burigana
\etals 1997, Maino \etals 1999; Keih\"anen \etals 2004; Efstathiou
2005). It is also seen in the $X$, $E$ and $B$ power spectra from
full-scale simulations of the polarised {\it Planck} $217$GHz
detectors (Ashdown \etals 2006).

For the knee-frequencies\footnote{The characteristic frequency above
which the detector noise is approximately white.} and noise levels
expected of {\it Planck} detectors, the noise power spectra for both
temperature and polarization will be accurately white-noise above
multipoles of $\ell \simgt 20$.  For the temperature power spectrum,
the effect of residual striping errors on the power spectrum at
multipoles $\ell \simlt 20$ should be negligible, and so it should be
an excellent approximation to treat the noise as pure
white-noise. However, the simulations of Ashdown \etals (2005) show
that the effects of striping errors on the $E$ and $B$-mode power
spectra at low multipoles will be comparable to (or will dominate) the
intrinsic signal. These errors should therefore be folded into
estimates of the power spectrum covariance matrices. This can be done
by evaluating the QML estimator on  low resolution maps, provided
one can adequately approximate the noise covariance matrices $N_{ij}$
for these low resolution maps. In Section 5 we showed that it is easy to
compute $N_{ij}$ for a low resolution map if the noise is strictly 
uncorrelated. However, the analogous problem for {\it Planck}-like
correlated noise has not yet been solved and is currently under
investigation. 

  Since the noise for a {\it Planck}-like experiment is accurately
white at some characteristic multipole, a hybrid estimator based on
PCL estimates at high multipoles and QML estimates at low multipoles,
is well motivated. Furthermore, by applying a QML estimator at low
multipoles, it is possible to eliminate almost completely the
troublesome effects of $E$ and $B$ mode mixing associated with sky
cuts. We have previously argued (E04, Efstathiou 2004b) that maximum
likelihood estimators should be used to determine the temperature
power spectrum at low multipoles on a cut sky to eliminate the large
`estimator-induced' variance inherent to PCL estimators. The problem
of estimator-induced variance is even more acute for PCL estimators of
the polarization power spectra ({\it cf} CC05). In particular, for the
PCL estimators discussed here, mode-mixing places significant limits
on the amplitude of the tensor-to-scalar ratio, $r$, that can be
probed even in a noise-free experiment. For example, the tests
described in Section 3.4 show that it is difficult to probe below $r
\sim 0.1$ using a PCL estimator applied to maps with a Kp0 sky-cut. As
Smith (2005) has shown, it is possible to define fast estimators based
on unambiguous $E$ and $B$ modes on a cut sky. However, applying a QML
estimator to low (or intermediate) resolution maps is also fast and is
very close to optimal. In fact, if one wants to incorporate QML power
spectrum estimation into a Monte-Carlo chain, then as equation
(\ref{MLN1}) shows, the time taken to evaluate a QML power spectrum is
no greater than the time taken to evaluate a fast spherical transform.

\vskip 0.1 truein

\noindent
{\bf Acknowledgements:} I thank members of the Cambridge Planck
Analysis Centre, especially Mark Ashdown and Anthony Challinor, for
helpful discussions.


\begin{thebibliography}{}

\bibitem[\protect\citename{AHS}2005]{AHS05}
Amarie M., Hirata C., Seljak U.,  2005, astro-ph/0508293.

\bibitem[\protect\citename{Aetal}2006]{Aetal06}
Ashdown M.A.J., {\it et al.}, 2006, in preparation.

\bibitem[\protect\citename{Betal03a}2003]{Betal03a}
Bennett, C. \etal, 2003a, ApJS, 148, 1.

\bibitem[\protect\citename{Betal03b}2003]{Betal03b}
Bennett, C. \etal, 2003b, ApJS, 148, 97.

\bibitem[\protect\citename{Betal96}1996]{Betal96}
Bersanelli M. \etal, 1996, COBRAS/SAMBA Report on the Phase A Study. ESTEC.



\bibitem[\protect\citename{Betal99}1999]{Betal99}  Bond J.R.,
Crittenden R.G.,  Jaffe A.H., Knox L., 1999,  Comput. Sci .Eng.,  1,  21.

\bibitem[\protect\citename{BJK98}1998]{BJK98}
Bond J.R., Jaffe A.H., Knox L., 1998, PRD, 57, 2117.

\bibitem[\protect\citename{B99b}1999]{B99b} Borrill J., 1999, in
EC-TMR Conf. Proc. 476, 3K Cosmology, eds. L. Maiani, F. Melchiorri
and N. Vittorio, Woodbury AIP, 277. 

\bibitem[\protect\citename{BCT05}2005]{BCT05}
Brown M.L., Castro P.G., Taylor A.N.,  2005, MNRAS, 360, 1262.

\bibitem[\protect\citename{B03}2003]{B03}
Bunn E.F., 2003, New Astron. Reviews, 47, 987.

\bibitem[\protect\citename{BZTO03}2003]{BZTO03}
Bunn E.F., Zaldarriaga M., Tegmark M., de Oliveira-Costa A., 2003,
PRD, 67, 023501.

\bibitem[\protect\citename{BMMDMBM97}1997]{BMMDMBM97}
Burigana C., Malaspina M., Mandolesi N., Danese L., Maino D., Bersanelli M.,
Maltoni M., 1997, Internal Report ITESRE. astro-ph/9906360.

\bibitem[\protect\citename{CC05}2005]{CC05}
Challinor A.D.,  Chon G., 2005,  MNRAS, 360, 509. (CC05)

\bibitem[\protect\citename{CCPHS04}2004]{CCPHS04}
Chon G., Challinor A., Prunet S., Hivon E., Szapudi I., MNRAS, 350, 914.

\bibitem[\protect\citename{DKP01}2001]{DKP01}
Dor\'e O., Knox L., Peel A., 2001, PRD, 64, 3001.

\bibitem[\protect\citename{E03a}2003]{E03a}
Efstathiou G., 2003,  MNRAS,  346, L26.

\bibitem[\protect\citename{E04}2004]{E04}
Efstathiou G., 2004a,  MNRAS, 349, 603. (E04)

\bibitem[\protect\citename{E04b}2004]{E04b}
Efstathiou G., 2004b,  MNRAS, 348, 885.

\bibitem[\protect\citename{E05}2005]{E05}
Efstathiou G., 2005,  MNRAS, 356, 1549.



\bibitem[\protect\citename{HB96}1996]{HB96}
Hamaker J.P., Bregman J.D., 1996, A\&AS, 117, 161.

\bibitem[\protect\citename{HG03}2003]{HG03}
Hansen F.K., G\'orski K.,  2003, MNRAS, 343, 559.

\bibitem[\protect\citename{Hetal03}2003]{Hetal03}
Hinshaw G. \etals  2003, ApjS, 148, 135.

\bibitem[\protect\citename{HW97}1997]{HW97}
Hu W.,  White M., 1997, New Astronomy, 2,  323.

\bibitem[\protect\citename{KKS97}1997]{KKS97}
Kamionkowski M., Kosowsky A., Stebbins A., 1997, PRD, 55, 7368.

\bibitem[\protect\citename{KKPMB04}2004]{KKPMB04}
Keih\"anen E., Kurki-Suonio H., Poutanen T., Maino, D., Burigana C., 
2004,  A\&A, 428, 287.


\bibitem[\protect\citename{Ketal03}2003]{Ketal03}
Kogut A., \etal, 2003, ApJS, 148, 161.

\bibitem[\protect\citename{Ketal02}2002]{Ketal02}
Kovac J. M., Leitch  E. M., Pryke C., 
Carlstrom J. E., Halverson N. W., Holzapfel W. L., 2002,
Nature, 420, 772.

\bibitem[\protect\citename{Letal02}2002]{Letal02} Leitch E. M. \etals 
Nature, 420, 772.

\bibitem[\protect\citename{L03}2003]{L03}
Lewis A., 2003, PRD, 68, 083509.

\bibitem[\protect\citename{LCT02}2002]{LCT02}
Lewis A., Challinor A., Turok N.,   2002, PRD, 65, 023505.

\bibitem[\protect\citename{L84}1984]{L84}
Lyth D.H., 1984, Phys Lett B., {\bf 147}, 403L.

\bibitem[\protect\citename{OSH99}1999]{OSH99}
Oh S.P., Spergel D.N., Hinshaw G., 1999,  ApJ, 510, 551.

\bibitem[\protect\citename{Metal99}1999]{Metal99}
Maino D., {\it et al.}, 1999,  A\&A Suppl. Ser., 140, 383.

\bibitem[\protect\citename{Metal05}2005]{Metals05}
Montroy T.E. \etal, 2005, astro-ph/0507514.

\bibitem[\protect\citename{P03}2003]{P03}
Pen U.L., 2003,  MNRAS, 346, 619.

\bibitem[\protect\citename{Retal04}2004]{Retal04}
Readhead A.C.S., \etals  2004, Science, {\bf 306}, 836.

\bibitem[\protect\citename{S05}2005]{S05}
Smith K. M.,  2005, submitted to PRD. astro-ph/0511629.

\bibitem[\protect\citename{Setal03}2003]{Setal03}
Spergel D.N., \etal,   2003, ApJS, 148, 175.

\bibitem[\protect\citename{S04}2004]{S04}
Stompor R., White M.,  2004, A\&A, 419, 783.

\bibitem[\protect\citename{T04}2004]{T04}
Taylor, A.C. \etal,   2004, {\it 
To appear in the proceedings of the XXXVIXth Rencontres de Moriond "Exploring the Universe}, 
astro-ph/0407148.



\bibitem[\protect\citename{T97}1997]{T97}
Tegmark M., 1997, PRD, 55, 5895.

\bibitem[\protect\citename{TdO97}2001]{TdO01}
Tegmark M., de Oliveira-Costa A.,  2001, PRD, 64, 063001. (TdO01)


\bibitem[\protect\citename{PC05}2005]{PC05}
The Planck Consortia, 2005,  {\it The Scientific Programme of Planck}, Efstathiou G., Lawrence C., Tauber J. (eds),
ESA-SCI(2005)-1, ESA Publications.


\bibitem[\protect\citename{VMK88}1988]{VMK88}
Varshalovich, D.A., Moskalev A.N., Khersonskii V.K., 1988, {\it Quantum Theory of
Angular Momentum}, World Scientific, Singapore.


\bibitem[\protect\citename{ZS97}1997]{ZS97}
Zaldarriaga M., Seljak U., 1997, PRD, 55, 1830.

\end{thebibliography}
\end{document}